\documentclass[a4paper,5p,fleqn]{cas-dc}
\usepackage[english]{babel}
\usepackage[authoryear]{natbib}
\usepackage[utf8]{inputenc}
\usepackage{graphicx}
\usepackage{epstopdf}
\usepackage{aas_macros}
\usepackage{color}
\usepackage{xcolor}
\usepackage{hyperref,lastpage}
\begin{document}

\let\WriteBookmarks\relax
\def\floatpagepagefraction{1}
\def\textpagefraction{.001}
\title{Estimation of Classical Cepheid's Physical Parameters from NIR Light Curves}
\shorttitle{Estimation of Classical Cepheid's...}
\shortauthors{Balázs L.G. et~al.}

\author[1,2]{Lajos G. Balázs}[]
\cormark[1]
\ead{balazs.lajos@csfk.org}

\author[1,2,3]{Gábor B. Kovács}[orcid=0000-0002-2265-3788]

\affiliation[1]{organization={HUN-REN CSFK Konkoly Observatory},
            addressline={Konkoly-Thege M. \'ut 13-17}, 
            city={Budapest},
            postcode={H-1121},
            country={Hungary}}
\affiliation[2]{organization={Department of Astronomy, Eötvös Loránd University},
            addressline={Pazmány Peter setány 1/A}, 
            city={Budapest},
            postcode={H-1117}, 
            country={Hungary}}
\affiliation[3]{organization={Gothard Astrophysical Observatory, Eötvös Loránd University},
            addressline={Szent Imre h. u. 112.}, 
            city={Szombathely},
            postcode={H-9700}, 
            country={Hungary}}

\begin{abstract}
Recent space-borne and ground-based observations provide photometric measurements as time series. The data points are nearly continuous over a limited observational interval or randomly scattered over a long period. 
The effect of interstellar dust extinction in the near-infrared range is only 10\% of that measured in the visual (V) range. However, the sensitivity of the light curve shape to the physical parameters in the near-infrared is significantly lower.
So, interpreting these types of data sets requires new approaches like the different large-scale surveys, which create similar problems with big data.\\
Using a selected data set, we provide a method for applying routines implemented in R to extract most information of measurements to determine physical parameters, which can also be used in automatic classification schemes and pipeline processing.\\
We made a multivariate classification of 131 Cepheid light curves (LC) in J,H, and K colors by applying routines of R, where all the LCs were represented in 20D  parameter space in these colors separately. Performing a Principal Component Analysis (PCA), we got an orthogonal coordinate system and squared Euclidean distances between LCs. The PCA resulted in 6 significant eigenvalues, which allowed us to reduce the 20-dimension to 6.  
We also estimated the optimal number of partitions of similar objects and obtained it equal to 7 in each color; their dependence on the period, absolute magnitude, amplitude, and metallicity are also discussed. We computed the Spearman rank correlations, and concerning periods and absolute magnitudes, the first three PCs had correlations at a very high significance level. Similar computations revealed significant relationships between the amplitude and the first two PCs, but the LCs depend only marginally on the metallicity in H and K colors.\\
The method shown can be generalized and implemented in unsupervised classification schemes and analysis of mixed and biased samples. The analysis of a sample of classical Cepheids observed only in near-infrared bands resulted in the information coded in the light curves being insufficient to determine the stars' metallicity and identified the mass as the dominating quantity to form the shape of LCs in our sample. 
\end{abstract}

\begin{highlights}
\item We made a multivariate classification of 131 Cepheid light curves in J,H, and K colors by applying routines of R.
\item A Principal Component Analysis was performed and determined the optimal number of similar object groups.
\item It was shown that the most important physical parameter influencing the shape of NIR light curves is the mass.
\item Our method can be generalized and implemented in unsupervised classification schemes and analysis of mixed and biased samples.
\end{highlights}

\maketitle
\section{Introduction}
The Cepheid-type pulsating variables have a fundamental role in numerous fields of astrophysics. The discovery of the period-luminosity relation by  \cite{Lea1912} led to the recognition of the existence of extra-galactic systems determining their distances and resulting finally in the Hubble law of the expanding Universe \citep{Hub1925,Hub1929}. The basic role of the period-luminosity relation in the cosmological distance ladder still motivates recent works to remove the possible systematic errors and reduce the stochastic scatter in this relation and to find the suitable photometric color range where these effects are minimal  \citep{Che2017,Nge2018}.

Already in the twenties of the last century \cite{Her1926} studied the LCs of these objects extensively and found a remarkable relation between the period and the form of the LCs secured in the optical range. He claims: 
\emph{'the most striking general feature is perhaps the relatively great symmetry of the light curves at periods from about $ 9^d$ to $13^d$ as compared with both shorter and longer ones; outside this region, the LCs display remarkable skewness - fast ascending accompanied with slower descending and sometimes exhibiting a bump on the descending branch'}.

Three fundamental physical parameters can be obtained from the multi-color photometric observations: the period, the absolute brightness or the distance - applying the period-luminosity relations, and the effective temperature - from calibrated color indexes.
Unfortunately, the metallicity cannot be obtained directly from the photometric data, so one needs spectroscopic observations and large-size telescopes
\citep{Gen2015,Inn2019,Luc2019}.  Notwithstanding, studies of the shape of the LCs have shown that it has some sensitivity to the metal content 
\citep{Kla2013,Bhar2018,Sin2019} and different formulae were developed to estimate its values from photometric data secured in shorter wavelengths without spectroscopic measurements. Similar and elaborated results were published before for RR Lyrae stars, and a general effort emerged to get methods to predict metallicity from the shape of LCs in the 90s. 

Considering the recent progress in these efforts, we can find solutions due to large-scale surveys that combine the photometric observations secured in different bands and merge those with spectroscopic ones, providing empirical database grids to estimate the physical parameters for millions of observed stars [see \citep{Sun2023} and references therein]. More precise and better-performing formulas can be found by selecting one group of variable stars, e.g., for RR Lyrae stars in \citep{Bhardwaj2023a}.

These methods work well only in those celestial areas where the interstellar extinction is mild and can be corrected for its effect straightforwardly; we can get data in shorter wavelengths and objects bright enough to get spectroscopic data. 
In dusty areas, the effect of interstellar extinction is proportional to $\lambda^{- \alpha}$ where $\lambda$ is the wavelength and $\alpha \approx 2$ \citep{Stead2009}. Due to this relationship, interstellar matter is more transparent about an order of magnitude in the near-infrared than in the visual spectral range. This property has an advantage in cosmological studies in correcting the effect of interstellar extinction. However, it showed a disadvantage in performing spectroscopic studies of the physical parameters in the NIR range, especially the metallicity, because a significant fraction of the characteristic spectral lines are in the visual spectral range.

Securing spectroscopic data in these areas is difficult and cannot provide data for numerous objects only for limited samples \citep{Matsunaga2023}. So, we cannot build up large experimental databases to estimate the parameters of newly observed samples similarly as it can be done otherwise. What we can do is map the knowledge compiled in other fields.

Focusing only on Cepheids, especially a distinguished class among the interesting objects embedded in dense, dusty environments like galactic disks, one should extract the most information from the available data. 
Hence, we need a well-determined, robust procedure to deduce all the available information from the measurements. Moreover, the data sets collected are usually biased and restricted, not covering the parameter space evenly and fully. At the same time, we should process the data set in quasi-real-time by pipelines without any supervision. 

\begin{figure}[h!]
  \includegraphics[width=0.92\linewidth]{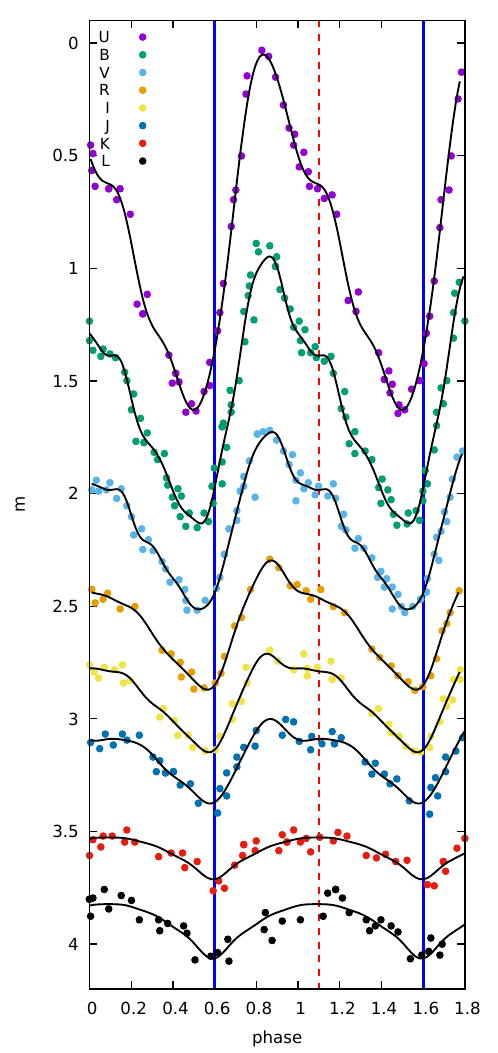}\\
  \caption{Typical light curves of a Cepheid ($\eta$ Aql) in different photometric bands; the data are taken from \citep{CepData68}}.\label{lctg}
\end{figure}

Browsing through the light curves of Cepheids secured in different photometric bands, we can find a simplification in the shape going into the infrared regime.
Figure \ref{lctg} shows light curves of a typical Cepheid ($\eta$ Aql) in a wide range over the observable spectrum; for a recent detailed analysis, see \cite{Nardetto23}. The striking general feature of the resemblance between the observations of Cepheids is the simplification of the light-curve structures and the phase shift toward longer wavelengths. 

Considering all these efforts, the general question may arise about how the form of the LCs depends on the basic physical parameters (e.g., mass, age,  rotation period, metallicity, etc.) in the different photometric bands. In other words, one can ask whether an inverse mapping exists from the observed time series of selected spectral ranges to the stellar parameters of radial pulsators. 

In addition to these questions, one can look for a robust and automatized unsupervised procedure to analyze not-comprehensive samples of objects piled up in large-scale surveys or independent projects. These limited samples usually cannot provide precise and well-determined detailed parameters but might help their classification and reveal unknown archeological features in the galactic structures.

Unfortunately, the theoretical side of the problem based on recent pulsation codes that treat the nonlinear dynamics of classical radial pulsators cannot directly provide the observable photometric light curves, only the bolometric one. However, static atmosphere models could be applied successfully to approximate the observed light variations in a large part of a pulsation cycle, but not at phases where the acceleration is high. The finer details of the variations, along with radial velocities, shocks, etc., cannot be modeled by recent one-dimensional codes that treated the atmosphere as a boundary condition. Although very intensive theoretical works are in progress to develop and manage multi-dimensional computations to treat the convective turbulence in the models and merge them with dynamic atmospheres, these are not available yet. So, we have to base our efforts on empirical relations and statistical studies.

A similar but different problem is the unsupervised classification of observed light curves. From a historical point of view, the Cepheids and W Virginis stars were considered to belong to the same class of variable stars, causing problems later. Hence, we should determine whether subclasses are possible in samples where the shape of the light curves seems similar in the first approximation, but subgroups can be separated. 

 An obvious question arises at this point. The light curves are the result of the pulsation dynamics operating inside the star. Can their shape be studied with the tools of multivariate statistics? The set of intensities resulted in sampling the light curves in given dates can be treated formally as coordinates of a vector in the parameter space. The number of these coordinate  values gives dimension of the parameter space.  A set of observed light curves represent a sample of these vector variables. 

Every coordinates of these vectors represent a scalar probability variable. Due to the internal dynamics of pulsation working behind the observed light curve these scalar variables obtained in this way are strongly correlated with each to the other. However, using PCA on their correlation matrix we can introduce uncorrelated vectors in parameter space defined above. Linear combination  of these orthogonal vectors fully reproduce those representing the LCs, including noise terms. In the following we studied the  statistical properties of the sample obtained by assuming the LCs as multidimensional vectors in the parameter space

The  R statistical package developed by \cite{Rref} contains a large number of libraries of routines usable for performing an analysis required by the mentioned efforts. These well-tested and sophisticated routines can provide all the data processing necessary for the abovementioned problems, and pipelines might quickly build from them. 
\footnote{The routines can also be called from Python codes.}
Nonetheless, they are less acquainted with the young generation of astronomers working in computerized data analysis. To advocate their application and bring their capabilities closer to real problems in this work, we demonstrate it on a step-by-step analysis considering a sample of  classical Cepheid variables but narrowing the applied photometric data to the near-infrared (NIR) bands. The applied routines can be packed into a compact program to provide a quick sample overview and find possible directions for further analysis. In our selected problem, we focused on Cepheids' NIR LCs and their metallicity, trying to decipher the metal content from LCs and compare those with spectroscopically determined values. 

Hence, we took a sample of  classical Cepheids covering a wide range of periods and analyzed their NIR LCs by  R routines to demonstrate their application in this type of problem.

The paper's organization is the following: Section 2 deals with the basic properties of Cepheid's near-infrared (NIR) LCs. Section 3 describes a classification procedure of LCs based on Principal Component Analysis (PCA). Section 4 discusses the main result and a possible relationship to the basic physical parameters, and Section 5 summarizes our conclusions.

\section{Characterization of light curves shapes}

After Hertzsprung's pioneering work on the shape of light curves of Cepheids, several authors developed quantitative methods to deduce physical parameters from the shape of observed asymmetric light curves in the optical bands. The LCs are periodic if one does not consider systems with a quasi-periodic behavior around this strictly periodic shape or contain low amplitude additional signals ($F_X$ modes). 

We can reckon that similar physical parameters generate similar LCs. Still, the opposite is not necessarily true, and the inversion of the LCs into physical parameters has to be studied in different colors separately. Hence, one may question the information in (J,H,K) bands in the near-infrared, where the LCs show fewer structural features than in the optical spectral range. 

Two methods in the literature are generally used to analyze periodic light curves: the decomposition into Fourier series and the Principal Component Analysis (PCA). 

\subsection{Fourier-series method}
 
Any periodic function can be approximated by a set of orthogonal functions defined over the period.  A prominent and mathematically well-determined way to study periodic functions is by the Fourier transformation or simply the Fourier series. Still, other sets of functions also can be used, e.g., the Chebhysev polynomials are better in numerous cases. 
 
However, the Fourier components are not the normal modes of the dynamical system oscillating in the fundamental or higher mode. 
Mathematically, the Fourier components give an equivalent representation of the function on a basic orthogonal system of functions, i.e., we mapped the continuous function onto a point of infinite dimensional space. Truncating the terms in the infinite series, we reduce the dimension to finite, assuming that the representation is still good enough. 

Considering a large enough sample of systems in the space of physical parameters, we get a scattered set of points drawing up a lower-dimensional geometric shape. One could get transformations simplifying the shape's topology and correlations with independently determined physical parameters by empirical (trial and error) methods.
 
The number of components of the Fourier series necessary to reproduce the LCs at a level of statistical goodness of fit depends on the photometric bands, and the distribution of data points beside unique features of LCs
\cite[in this context, see:][]{Nge2003,Nge2004,Nge2006,Ngo2006,Bha2015,
Bha2017}. However, due to Gibb's oscillation, the sharp features require many more terms, or one must use a weighting like Lanczos' method \citep{Lanczos66}.

\subsection{Principal Component Analysis (PCA)}

Since the pioneering work of \cite{Dee1964} on PCA, it has been widely used in diverse astrophysical problems. Because the PCA is related to the observed variables, it requires the lowest number of parameters to fit the observed values within the limit of statistical inference.

We can usually measure several quantities characterizing the observed system, but the values of parameters determining the system's behavior might be hidden from us. The basic idea behind PCA is to consider the measured observable values as  mutually dependent statistical variables and look for noncorrelated stochastic variables whose functions are observed.
The measurements might happen simultaneously or in different phases in a periodically variable process, as we observe in the case of Cepheids. {As we already mentioned above,} if we have an observation in $N$ phases, we can represent the continuous LC with an N-dimensional stochastic vector variable. When the number of independent physical parameters is lower than $N$, PCA gives the number of those and an orthogonal base for the subspace deduced from the statistical features of measurements. In this sense, the PCA gives well-determined estimates of the information in the observed data set and its lowest dimension interpretation.

The observations may refer to one system over many periods or similar systems over a limited time interval as a sample. In the first case, any instability in the system will appear as some stochastic part. The PCA can estimate the expected values at a given phase; the residuals characterize the randomness around this (e.g., Blazhko effect at RR Lyrae stars or granulation noise in Cepheids). When the data refers to a type of object covering a parameter space, giving a well-distributed sample of the studied systems, the PCA can span the subspace of substantial parameters. Its dimension might be higher than the independent parameters, but it allows embedding all of the systems into it while removing the redundant dimensions. However, if the sample distribution is biased, we cannot perceive all of the parameters, only those represented well in the sample.

\section{The analyzed data sets}

To demonstrate the procedure, we used a  database of 131 LCs of \cite{Mon2011} in our analysis. The apparent J,H,K colors of our program stars are given by  \citet{Mon2011}, and their parallaxes are available in the Gaia DR2 archive \citep{Lur2018}. We corrected the interstellar extinction by comparing the observed J-H, and K-H color indices with their intrinsic values given by \citet{Bes1988}. The ratios between total and selective extinction were also taken from this paper. The period and metallicity data are taken from \citep{Bed2000,Gro2018}, respectively.

The presented PCA method can be automatized and pipelined, providing a possible way for unsupervised classification tasks.

\section{PCA of light curves}
\label{pca}

The application of PCA to the Cepheid LCs \cite{Kan2002} showed beforehand that it could be used effectively in this context and concluded that this approach is more efficient than Fourier analysis for finding changes in the LC as a function of physical quantities (e.g., period). Scrutinizing a high number (17608) of LCs \cite{Deb2009} demonstrated that, in general, the first few principal components (PCs) are enough to reconstruct the original light curves within the limits of the statistical inference. They also pointed out that the PCA technique can be used to classify variables into different variability classes in an automated, unsupervised way.

The PCA-based template fitting approach has several advantages, such as accurate parameter estimation, simultaneous fitting to multiple passbands, and estimate of the mean metallicities of Cepheid samples \citep{Tan2005}.

Nevertheless, in a practical approach, applying routines of well-tested software libraries is a great adventure. In the following, to advocate and help the application of PCA, we show a step-by-step method of applying the R statistical package and comment on its steps.

We used as a sample the NIR light curves of \cite{Mon2011} containing
131 fundamental mode Cepheids. Because the individual Cepheids have different periods, the LCs have to be represented depending on phases instead of the time series of observations to make them comparable  \cite[for the details of the phasing, see, e.g.][]{Inn2015}. In addition, the curves are normalized in amplitude.

The targets were observed in a finite set of discrete moments, resulting in 22 points on average. Fitting these observations by the \textcolor{red}{smooth.spline()} and subsequently applying the \textcolor{red}{predict()} procedures of the R statistical package\footnote{http://cran.r-project.org}, we estimated the brightness in 20 points, distributed uniformly in phase. After this standardization, we transformed the LCs into the 20-dimensional parameter space where every LC is  represented as a point.

\subsubsection{Comparison of shapes}

We may characterize the similarity between any two
LCs by the $\chi^2$ distance as, 

\begin{equation}\label{chi2}
    \chi^2_p= \sum^p_{i=1}\frac{(x^k_i-x^l_i)^2}{\sigma^2_i}
\end{equation}

\noindent where $x$ is a representation of an LC as a $p$-dimensional random variable with $\{x_1, x_2, \ldots , x_p\}$ components, representing the
brightness at a given color and phase. $\sigma^2_i$ is the
variance along the $i^{th}$ dimension, $p$ is the dimension of the
parameter space, $k$ and $l$ are  indexes of two arbitrarily
chosen objects from the sample of the LCs.

Equation (\ref{chi2}) formally defines a  variable of $p$ degrees
of freedom. However, it is not the sum of squares
of independent variables because they may correlate. As a consequence, a change along an $x_i$ component of $x$
may indicate a change along another $x_j$, due to their correlation.   One gets uncorrelated variables by solving the following eigenvalue problem \cite[details of this procedure can be found in the paper of ][]{Deb2009}:

\begin{equation}\label{eigen}
    R_{ij}a_j=\lambda a_i
\end{equation}

\noindent where $R_{lk}$ is the correlation matrix between the
$\{1,2, \ldots ,p \}$ components of the $x$ variable, $a_i$ is an
eigenvector of unite length $\{a^2_1+a^2_2+ \ldots + a^2_p=1\}$
 and $\lambda$ is the eigenvalue.  The PCs are uncorrelated and completely
reproduce the original LCs, including the noise term.

Although the PCs reproduce the observed LCs completely, their basic shape is determined by a much smaller number of physical parameters (e.g., mass, age,
metallicity). Therefore, one expects fewer PCs necessary to reproduce the basic characteristics of the LCs. 

To perform the above-described procedure, we used the \textcolor{red}{\tt princomp()} procedure in the $stats$ library of the  R statistical package.

The magnitudes of the $\lambda$ eigenvalues give information on
the importance of the given PC in reproducing the original
observed variables and give the variances of the PCs. Ordering $\lambda$s in descending order and evaluating the following formula,  one gets the cumulative  
percentage of the total sum of variances expressed by the first $m<p$ PCs:

\begin{equation}\label{lvar}
   Cum. prop.= \frac{\sum\limits^m_{i=1} \lambda_i}{\sum^p_{i=1}\lambda_i}
\end{equation}

\begin{figure}
  \includegraphics[width=8cm]{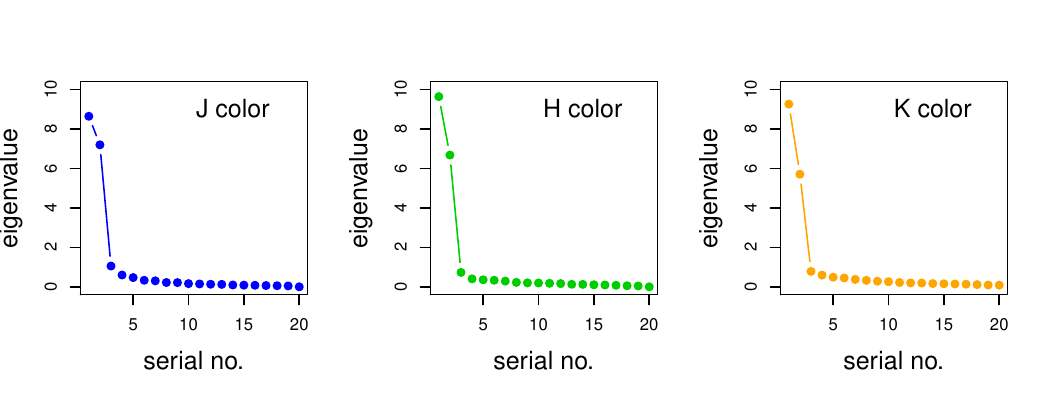}\\
  \caption{Eigenvalues resulted in the PCA in descending order. The first two eigenvalues
   are followed by a rapid value decrease with an inflection at about 5.
   The first 6 eigenvalues are kept for further study.}\label{eig}
\end{figure}

\begin{table}
\begin{tabular}{|c|cccccc|}
  \hline
  Color & PC1 & PC2 & PC3 & PC4 & PC5 & PC6 \\
  \hline
  J & 0.432 & 0.792 & 0.844 & 0.874 & 0.898 & 0.914 \\
 H & 0.482 & 0.816 & 0.852 & 0.872 & 0.890 & 0.907 \\
 K & 0.463 & 0.748 & 0.787 & 0.817 &  0.841 & 0.864 \\
  \hline
\end{tabular}
\caption{Cumulative proportions expressed by the fraction of sum
of the first $m < p$ and the total number of PCs (see Eq.
(\ref{lvar})). The $\lambda$ eigenvalues are in descending order.
We took into account the first 6 PCs, expressing about 90\% of the
total variance.}\label{cump}
\end{table}

\begin{figure}
  \includegraphics[width=8cm]{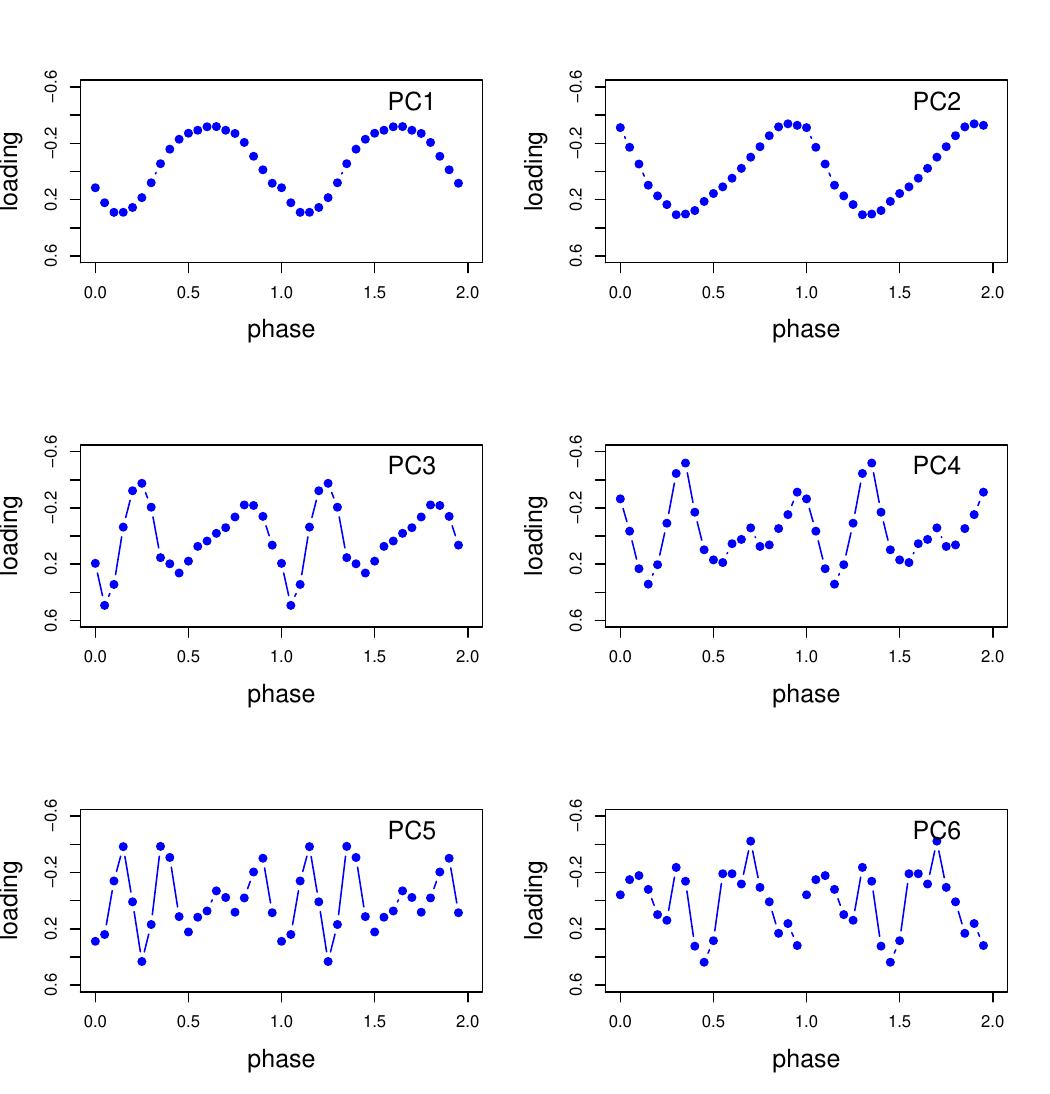}\\
  \caption{Elements of the eigenvectors (loadings) displayed as
the function of the phase. They can be considered as templates for
reproducing the observed LCs in the J color as linear
combinations.}\label{Jpl}
\end{figure}

\begin{figure}
  \includegraphics[width=8cm]{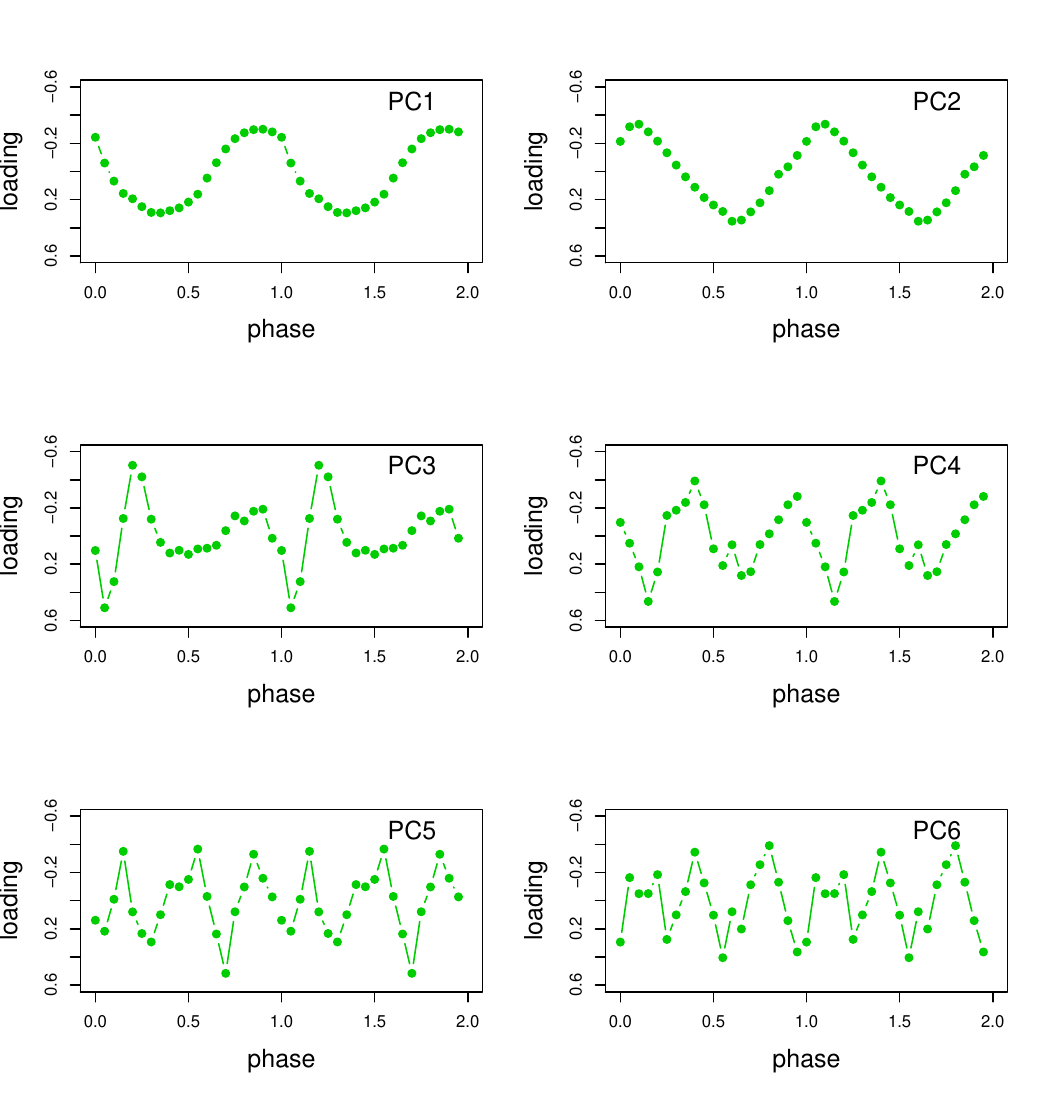}\\
  \caption{The same as Figure \ref{Jpl} but in the H color}\label{Hpl}
\end{figure}

\begin{figure}
  \includegraphics[width=8cm]{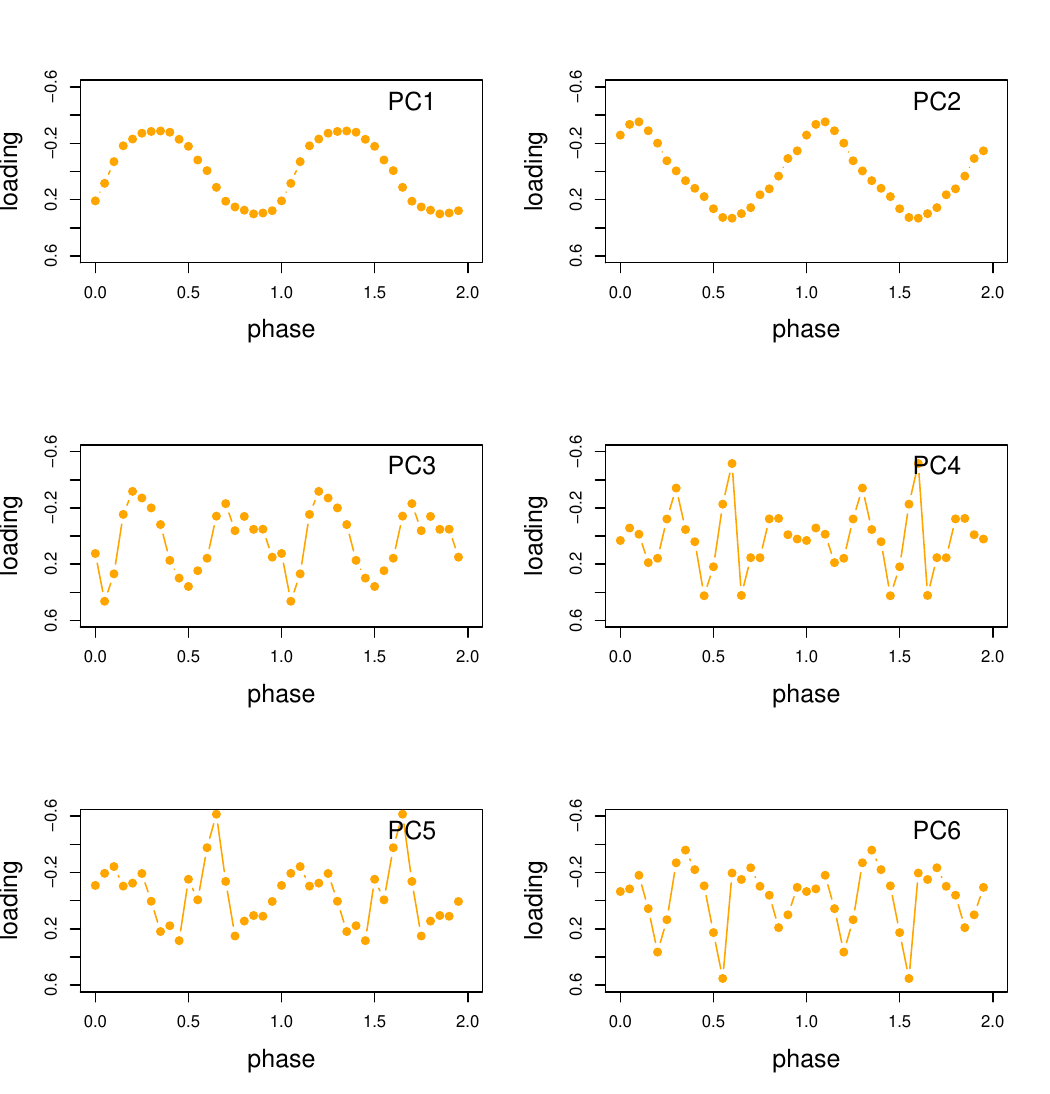}\\
  \caption{The same as Figure \ref{Jpl} but in the K color}\label{Kpl}
\end{figure}

\noindent The values of the cumulative proportions are given in
Table~\ref{cump} and the eigenvectors belonging to $\lambda$s are
templates for reproducing the observed LCs. We calculated them
 in each color separately and displayed them in Figure \ref{Jpl}, \ref{Hpl}, and \ref{Kpl}.

As the eigenvectors are orthogonal to each other, considering the first 6 of them as an
orthogonal base in the parameter space of 6-dimensions, one can
approximate the LCs of the sample by their linear combination.

\subsection{Partitioning the parameter space}
\label{medoid}

Taking the first 6 PCs, we get an orthogonal
basis in the parameter space, and we can define the similarity between the approximated LCs likewise to Equation (\ref{chi2}):

\begin{equation}\label{pdist}
     \chi^2_6= \sum^6_{i=1}\frac{(u^k_i-u^l_i)^2}{\lambda_i}
\end{equation}

\noindent where $u$ means the value (score) of the $PC$ and the
$\lambda$ eigenvalue gives its variance. The $\chi^2_6$ variable in
Equation (\ref{pdist}) can be used for partitioning the parameter
space into regions having LCs of some level of similarity.
Contrary to Equation (\ref{chi2}), however, the components of the
sum are not correlated, so a change in one of them does not have
any influence on others, and one can study its influence on the shape of the LC directly.

Since the  PCs are not correlated, their influence on the shape of
the LCs can be studied separately, and the dependence of the shape of
the LCs on the physical parameters can be studied through the
PCs' dependence on them. One may expect that similar physical parameters mean
similar PCs and, through them, similar LCs, but the opposite is not
necessarily true. Similar LCs do not indicate similar physical
parameters because a particular PC does not necessarily depend on a
particular physical parameter.

Based on the similarity defined by Equation (\ref{pdist}), we
partitioned the parameter space into regions of similar LCs,
according to this measure. There are several approaches to
partition the parameter space; a widely used method is the {\it
k-means clustering}.

Let's assume a priori the {\it k} number of the partitions and
start from initially arbitrarily defined centers of the
partitions in the parameter space. Then, one has to compute the distances of
all sample members from all the centers and assign all of them to the center of the closest one. After computing the
means of the coordinates within each partition, one has to replace the
originally chosen centers with the newly obtained means and repeat
the procedure of computing the distances from the new centers.
This  procedure \cite[k-means clustering,][]{Mac1967} is repeated
until reaching some convergence of the means.

Performing this procedure, the finally obtained centers do not
necessarily correspond to any objects in the sample. However, there is a
 method similar to the previous one, called {\it partitioning around
the medoid} (see textbook of \cite{Kau1990}) resulting in centers corresponding to one of the
objects, the {\it medoids} of the subgroup of similar objects. This method selects typical objects from the sample, and it is available in the {\tt
\textcolor{red}{pamk()}}  procedure of the {\it fpc} library of
 R. 

\begin{figure}
  \includegraphics[width=8cm]{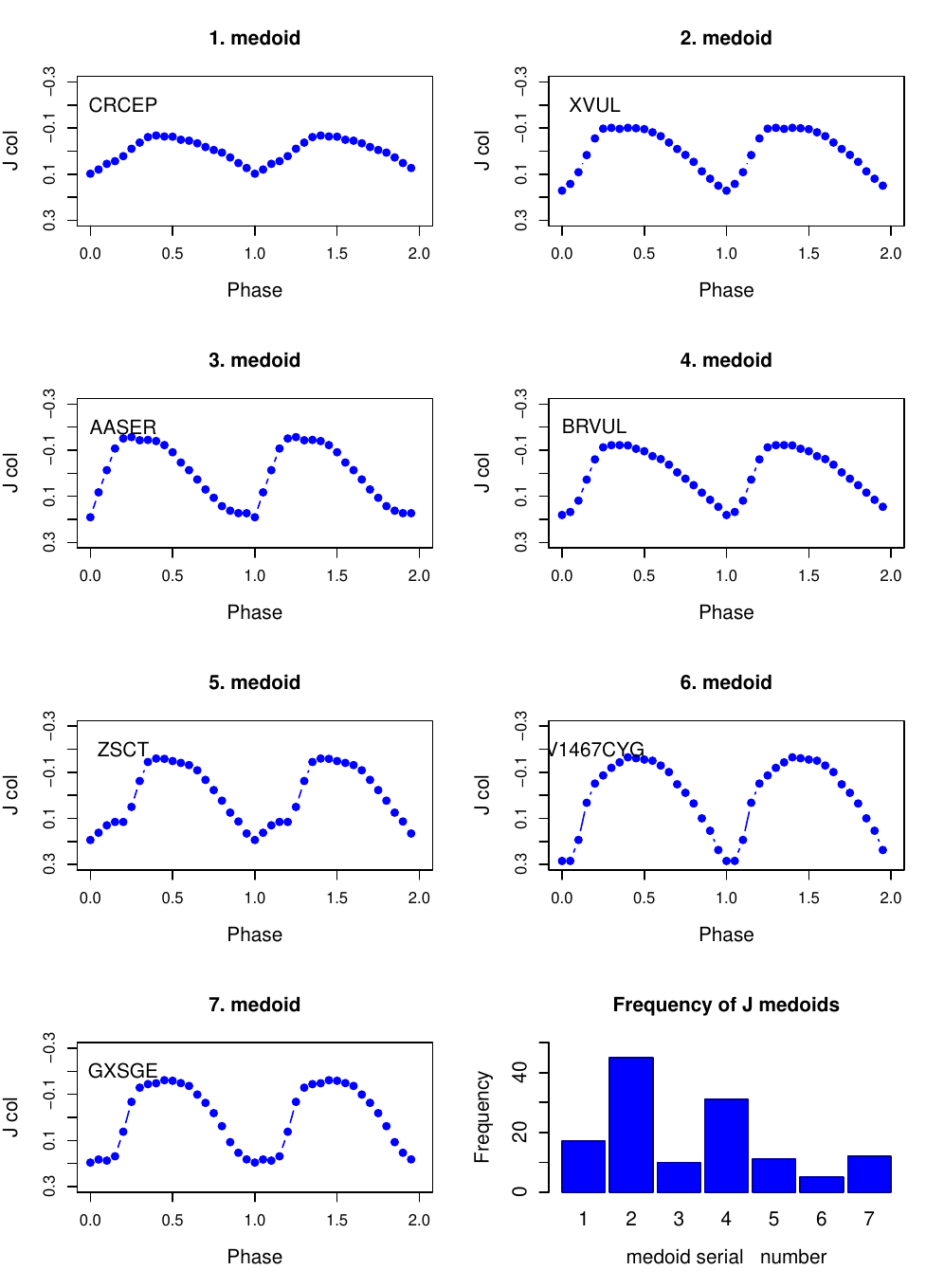}\\
  \caption{Typical LCs (medoids) in the J color.  The frequency of medoids
   is given in the bar chart at the bottom right. Note that the clustering algorithm somewhat arbitrarily gives the serial number of a medoid, and it does not necessarily have any further physical meaning. (The LCs were reconstructed
  from the first six PCs).}\label{Jmed}
\end{figure}

\begin{figure}
  \includegraphics[width=8cm]{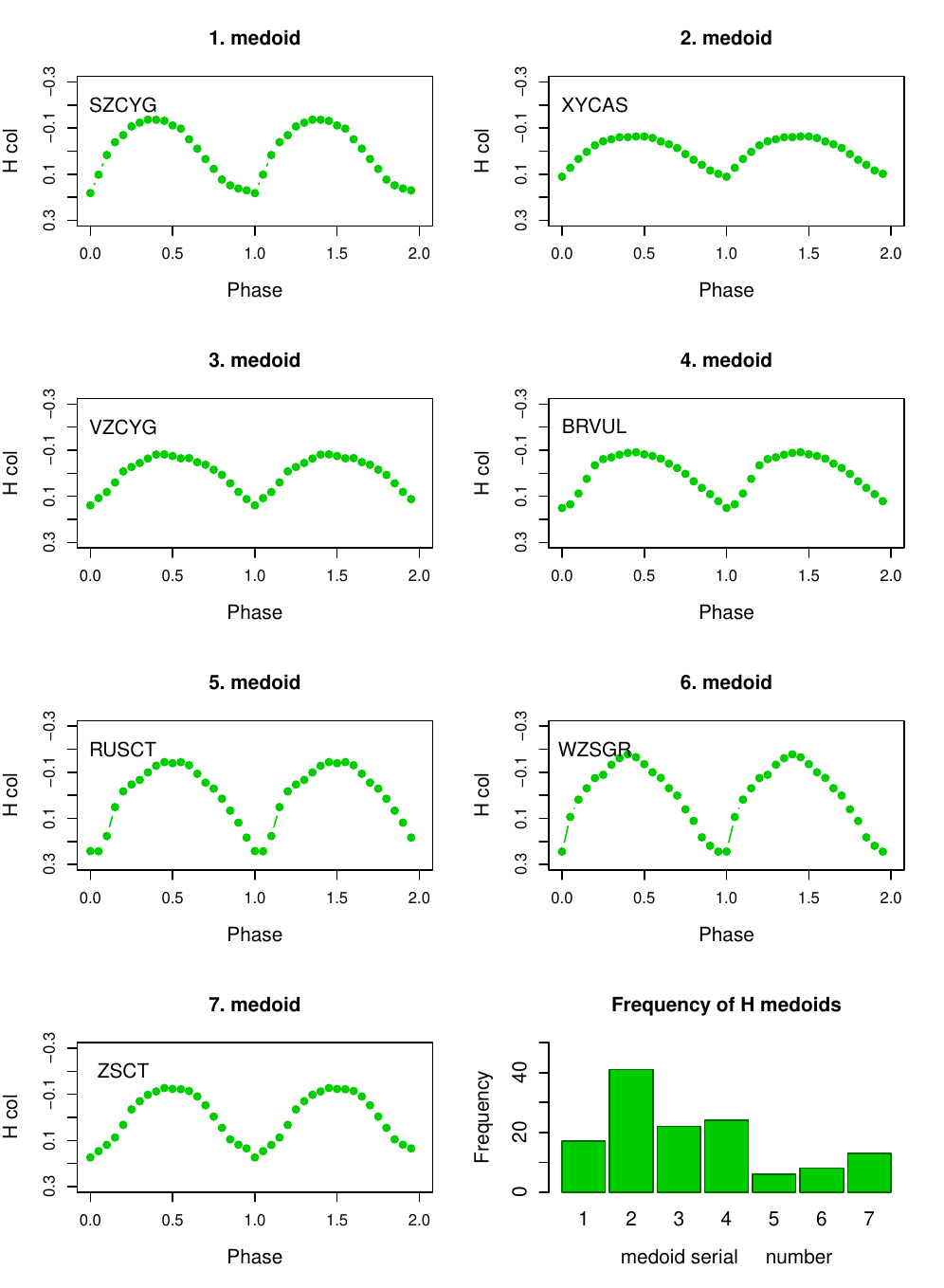}\\
  \caption{Typical LCs (medoids) in the H color. The frequency of medoids
   is given in the bar chart at the bottom right. (The LCs were reconstructed
  from the first six PCs).}\label{Hmed}
\end{figure}

\begin{figure}
  \includegraphics[width=8cm]{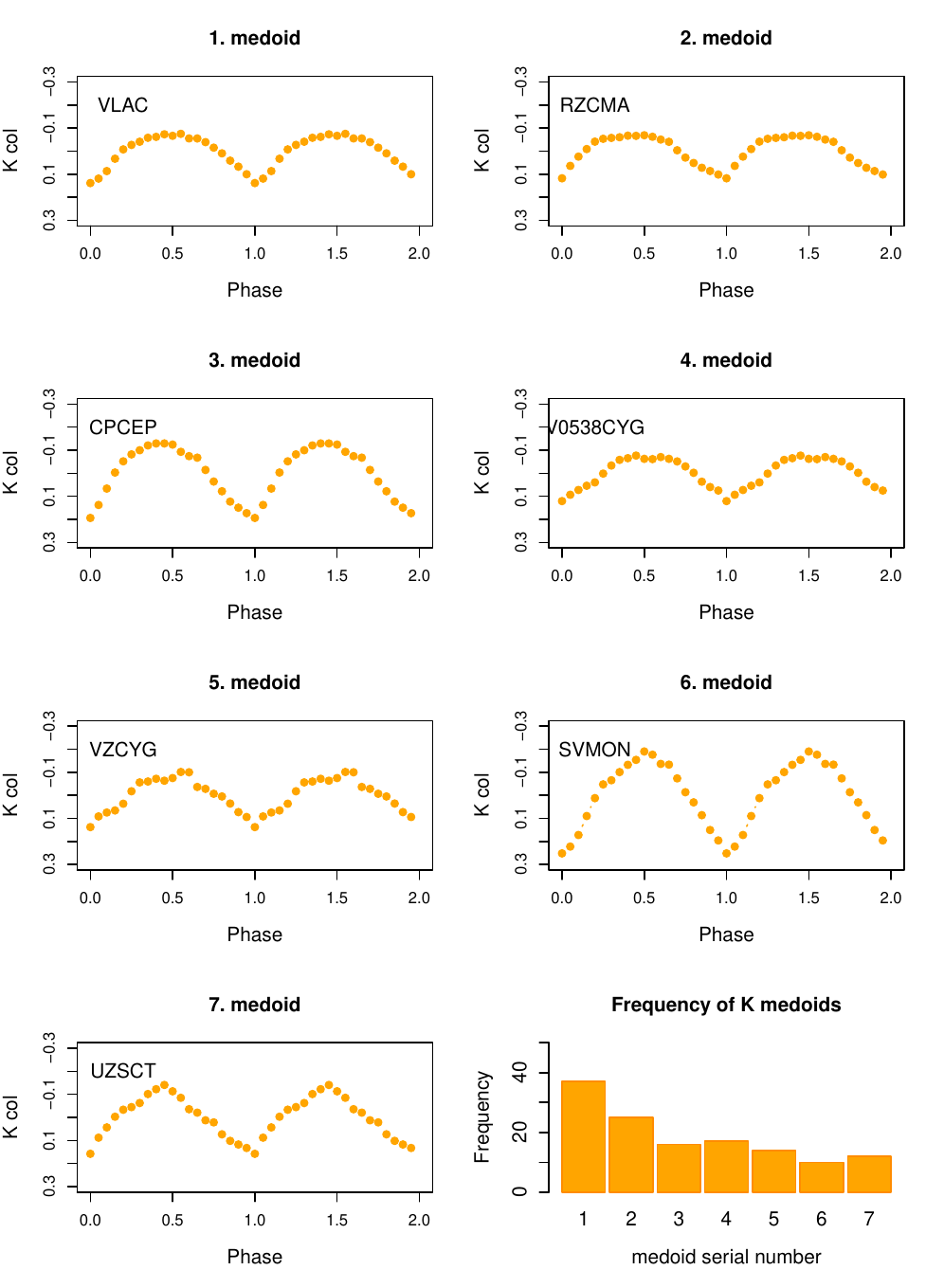}\\
  \caption{Typical LCs (medoids) in the K color. The frequency of medoids
   is given in the bar chart at the bottom right. (The LCs were reconstructed
  from the first six PCs).}\label{Kmed}
\end{figure}

We used this procedure to obtain template LCs.
 The \textcolor{red}{pamk()} procedure resulted in 7
groups in all of the $J, H$ and $K$ colors. The most typical LCs (the
medoids) are displayed in Figures \ref{Jmed}, \ref{Hmed}, and
\ref{Kmed} in all groups and in all colors.

The \textcolor{red}{pamk ()} procedure determines the optimal $k$ group number by making use of the {\it silhouette} method. 
This procedure determines the $a(i)$ average of the distances of a sample element measured to the other elements and the $b(i)$ distance of this element to the nearest group for each element of the sample.
The $s[i]$ silhouette of element $i$ is defined by the formula as $s[i] = (b[i] -a[i]) / MAX (b[i],a[i])$, where $MAX$ indicates the larger of $b[i], a[i]$. When grouping with different $k$ values, the optimal $k$ group number is at which the mean of $s[i]$ is maximal.

Regarding the optimal number of partitions obtained, we may ask whether it is only a characteristic of the given sample or general for similar samples. We used the {\it jackknife} resampling technique to answer this question. It obtains new samples by randomly omitting one of the observations and repeating the statistics performed on the original sample.
	
Using \textcolor{red}{jackknife()} procedure in the $bootstrap$ library of R,  we estimated the number of optimal partitions for each subsample. In this way, we obtained 131 new samples and repeated the partitioning algorithm on each. Table \ref{jack} shows the distribution of an optimal number of partitions in all colors. The most frequent number of partitions is 7, also obtained for our original LC sample. 

\begin{table}
	\centering
	\begin{tabular}{|r|rrrrr|r|}
		\hline
		{ No. groups} & 5 & 6 & 7 & 8 & 9 & 10 \\
		\hline
		{ J color} & 3 & 6 & { 107} & 13 & 2 & - \\ 
		{ H color} & 1 & 36 & { 74} & 12 & 4 & 4 \\ 
		{ K color} & 2 & 5 & { 64} & 13 & 4 & 39 \\ 
		\hline
	\end{tabular}
\caption{Results of the {\it jackknife} resampling. The most frequent number of optimum partitions is 7 in all colors}
\label{jack}
\end{table}

Since we performed the clustering algorithm separately in each
color, a certain medoid does not necessarily correspond to the same star in another color.
It is also the case concerning the serial number of the groups.
Due to this circumstance, one cannot make a one-to-one
correspondence between medoids in different colors. However, it is
evident already at first glance (see Figures
\ref{Jmed},\ref{Hmed} and \ref{Kmed}) that the medoids in the J
color are much more asymmetric than in  H and K, with respect to
the maximum of LCs. We will discuss the dependence of the LCs and
their medoids on the physical parameters in section \ref{disc}.

\section{Discussion}
\label{disc}

The classification of the LCs proceeded by using the distance
measure defined in Equation (\ref{pdist}). This approach gives
equal weight to all the involved PCs in the analysis. It could
happen that a physical quantity influences the PCs, contributing
to $\chi^2_6$ at the left side of Equation (\ref{pdist}), with
different strengths and, consequently, the final similarity between
LCs. In the following, we will discuss this effect in more detail.

\subsection{Dependence of the PCs on the period.}
\label{PCPer}
 In Subsection \ref{medoid}, we used 6 PCs to
reproduce the LCs and compute the mutual similarities. To see
the connection between the PCs and the period, we computed the
non-parametric Spearman's rank correlations. The results for the
first four PCs are given in Table \ref{Ppc} and note that PC5 and PC6 did not
give significant correlations.

\begin{table}
  \centering
  \begin{tabular}{|c|c|c|c|c|c|c|}
    \hline
   Color    &    PC1     &    PC2     &    PC3     &   PC4     \\
\hline
J: p-val.  &  < 1.0e-04  &  1.31e-03  &  < 1.0e-04  &  0.006   \\
H: p-val.  &  < 1.0e-04  &  < 1.0e-04  &  < 1.0e-04  & 0.77    \\
K: p-val.  &  < 1.0e-04  &  < 1.0e-04  &  < 1.0e-04  & 0.95    \\
\hline
  \end{tabular}
  \caption{PC - Period correlations obtained from Spearman's rank correlation
  to avoid the possible effect of outliers. Significant correlations are marked with boldface.
  The first three PCs have a very strong correlation in all of the colors.}
  \label{Ppc}
\end{table}

Figure \ref{eig} shows two strong eigenvalues in all three colors. The LC is a linear combination of the
eigenvectors (loadings) obtained from the PCA, and the PC scores
are the coefficients. As one can see in Figures \ref{Jpl},
\ref{Hpl} and \ref{Kpl} the first two eigenvectors (PC1 and PC2)
have a form similar to two orthogonal periodic functions with the period
of the LC but with a phase shift relative to each other. PC1 and PC2 are particularly important in reconstructing light curves using PCs.

\begin{figure}
  \centering
  \includegraphics[width=8cm]{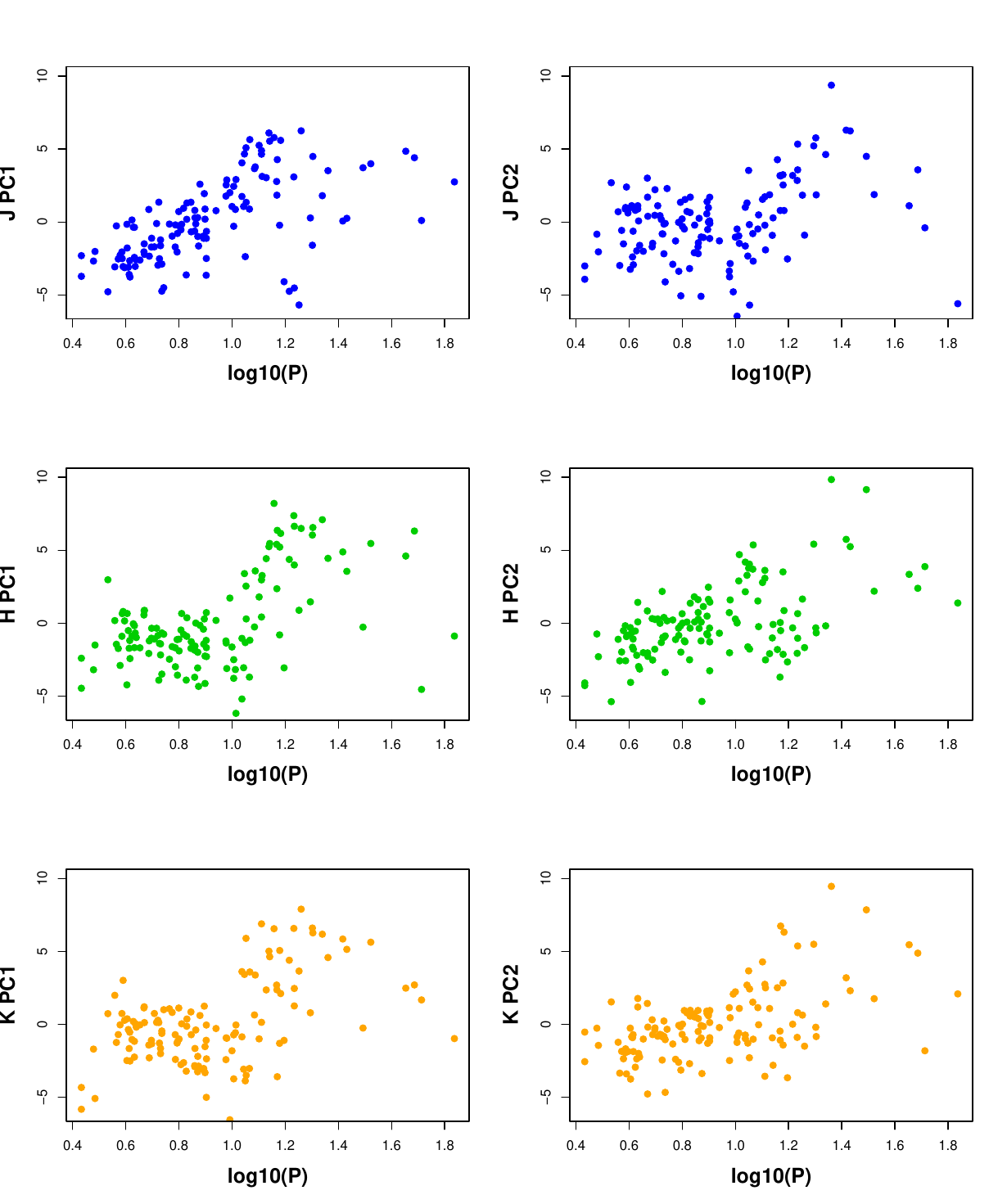}\\
  \caption{Dependence of the LC PC1 and PC2 principal components on the period in
  J,H, and K colors.
  These two PCs have the largest contribution to shaping
  the LCs. See the drastic change in the distribution of PC1 values over the
  period of 10 days, in particular in the H and K colors. This period is
  also critical for the amplitude-period relationships in the J, H, and K
  colors. (See Figure \ref{JHKPAmp}).}\label{JHKPhP}
\end{figure}

Figure \ref{JHKPhP} shows at first glance that the points distribution differs significantly in the $P \, < \, 10 \, days$ and $P \, > \, 10 \, days$ range in all of the colors. \citet{Deb2009} obtained similar results for PC1 and PC2 from PCA of LCs in V and I colors. The phenomenon also occurs in the case of some LCs Fourier components dependence on the $P$ period \citep{Andr1988,Deb2009,Nge2003,Sim1981}. The phenomenon is caused by the$ P_2 / P_0 = 0.5$ resonance at $P = 10 \,  days$ period \citep{Sim1981}.

According to the $P$ {\it period} $\propto$ $\bar{\varrho}$ {\it
average density} relation  $P \sqrt{\bar{ \varrho}}= const$
\citep[discovered theoreetically  by ][]{Edd1926}. Accordingly,
a higher period means lower average density. A specified period
 means a specified density for the pulsation characteristics and, consequently, for the
NIR LC of the star.

Since the $\bar{\varrho}$ average density can be expressed by the
$M$ mass and $R$ radius of the star as $\bar{\varrho}=M/R^3$ it
gives a canonical scaling for the $P$ period as $P \propto R^{1.5}
M^{-0.5}$. \citet{Bon2001}, however, gives $logP = 1.86logR-0.8logM-1.7$
 yielding $P \propto R^{1.8} M^{-0.8}$ which, as they claim, does not differ significantly from
the canonical relationship in the $P < 10^d$ range but differs of
the order of 25\%-30\% for $P = 60^d$.

 Nevertheless, \citet{Gie1989} gives a relationship between $P$ and $R$ in the form of $logR=1.068+0.767logP$. Combining this equation with the relationship given by \citet{Bon2001} we get $P \propto M^\alpha$ where $\alpha=2.1$. (The same procedure yields $\alpha=3.3$ for canonical scaling).

We may conclude that the mass is a major parameter in the two strongest PCs, consequently shaping the LCs in our sample. This is an example of how one can reveal a relation between a physical parameter and the shape of a light curve. Of course, this is not a new insight into the physics of Cepheids but is used only to demonstrate the applicability of PCA to recover hidden relations.

\subsection{Dependence of the PCs on the absolute brightness}
\label{PCMJHK}
In addition to the period, absolute brightness is central to cosmology due to the period-luminosity relation. The distance to the star must be known and corrected for interstellar extinction to get the absolute brightness from the apparent one.

\begin{figure}
    \centering
    \includegraphics[width=8cm]{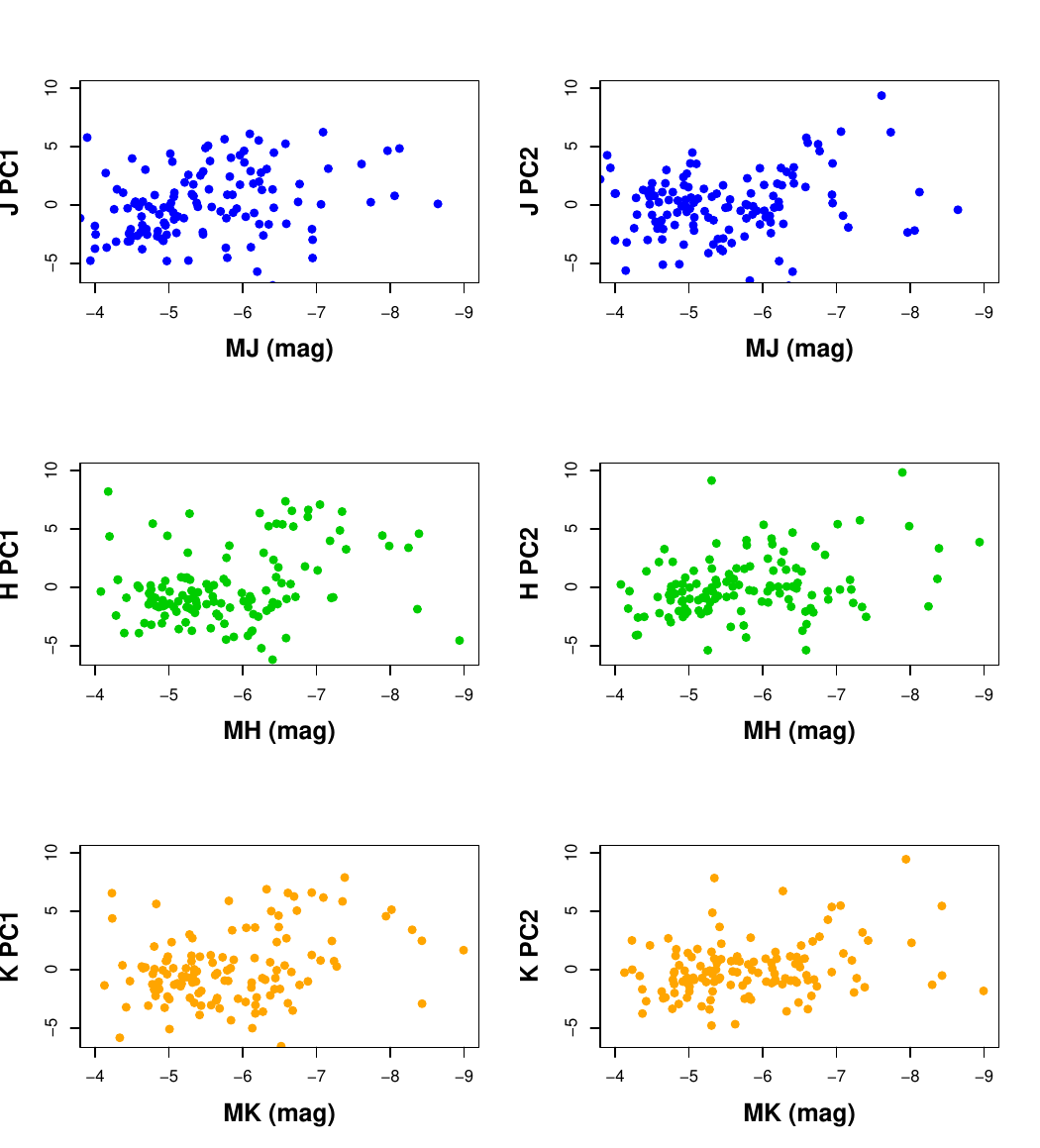}\\
    \caption{Dependence of the LC PC1 and PC2 principal components on the absolute magnitudes in J,H, and K colors. Note the drastic change of the point pattern at about -6 Mag. in H and K colors.}
    \label{MJHK_PC}
\end{figure}

The relationship between PC1, PC2, and absolute magnitudes in J,H,K colors are displayed in Figure \ref{MJHK_PC}. A characteristic feature of these figures is the change in the distribution of points at a given absolute brightness, especially in the H and K colors. 

\begin{figure}
    \centering
    \includegraphics[width=8cm]{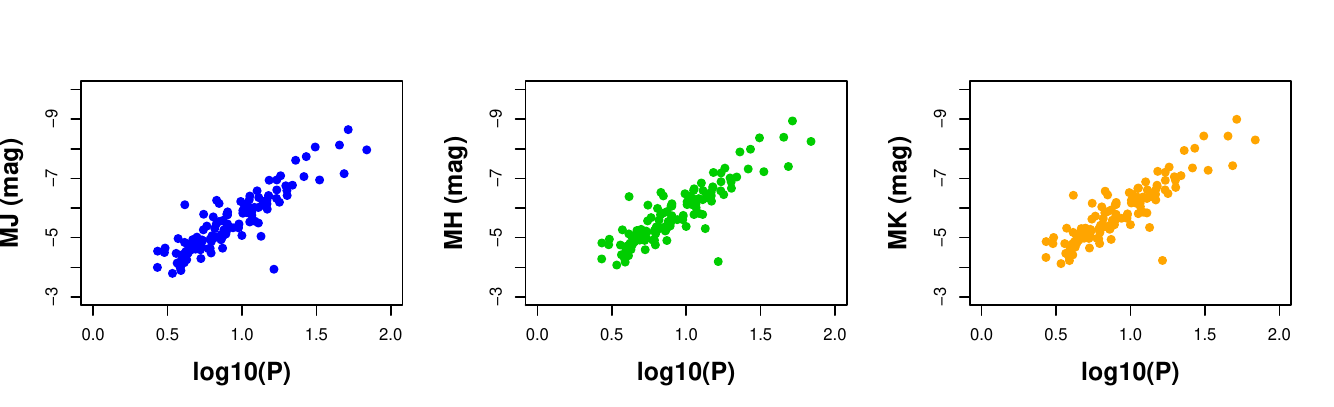}\\
    \caption{Period - absolute magnitude relationships in J,H, and K colors. The close linear correlation between these two quantities is fundamental in setting up the cosmological distance scale. }
    \label{MJHK_P}
\end{figure}

As one can infer from Figure \ref{MJHK_P}, these absolute magnitude values correspond to the $P=10^d$ characteristic transit period discussed in the previous subsection. However, it is worth mentioning the very close relationship between the absolute magnitude and period does not show any similar changes, as seen in Figure \ref{MJHK_PC} when passing the $P=10^d$.

\begin{table}
    \centering
    \begin{tabular}{|c|c|c|c|c|c|c|}
        \hline
        color    & PC1      & PC2      & PC3     & PC4 \\
        \hline
        MJ: pval &  < 1.0e-04 &  9.66e-04 &  1.43e-04 &  0.01  \\
        MH: pval &  < 1.0e-04 &  < 1.0e-04 &  1.96e-03 & 0.43  \\
        MK: pval &  6.91e-04 &  < 1.0e-04 &  3.06e-03 & 0.35 \\
        \hline
    \end{tabular}
 \caption{PC - absolute magnitude correlations obtained from Spearman's rank correlation to avoid the possible effect of outliers. Significant correlations are marked with boldface. The first three PCs have a very strong correlation in all of the colors.}
\label{MJHKpc}
\end{table}

Table \ref{MJHKpc}. shows the probability of purely random correlations between the PCs and absolute magnitude. Seemingly, PC1, PC2, and PC3 principal components have extremely tight correlations with absolute magnitude in all of the J,H,K colors. 

These tight correlations refer to a background variable on which they both depend very closely. Repeating the sequence of reasoning made in the previous subsection, we can identify this variable as the mass of the star.

\subsection{Dependence of the PCs on the LC amplitude.}
\label{PCAmp}

\begin{table}
  \centering
  \begin{tabular}{|c|c|c|c|c|c|c|}
    \hline
   Color    &    PC1     &    PC2     &    PC3     &   PC4    \\
  \hline
  Jamp: p-val.  &  3.08e-04  &  <  1.0e-04  & 0.047      & 0.311     \\
  Hamp: p-val.  &  < 1.0e-04  & < 1.0e-04 & 0.332    &  0.743   \\
  Kamp: pval.   &  < 1.0e-04  &  < 1.0e-04 & 0.236    & 0.910     \\
  \hline
  \end{tabular}
  \caption{PC - Amplitude correlations obtained from Spearman's rank correlation,
  to avoid the possible effect of outliers. Significant correlations are marked with boldface.
  The first two PCs have very strong correlation in all of the colors.}\label{PC_Amp}
\end{table}

Table \ref{PC_Amp} shows the Spearman's correlations of the first
four PCs on the LC amplitude in J,H, and K colors, characterizing
the strength of the dependence between these variables. The first
two PC have very significant correlations in all three
colors.

\begin{figure}
  \centering
  \includegraphics[width=8cm]{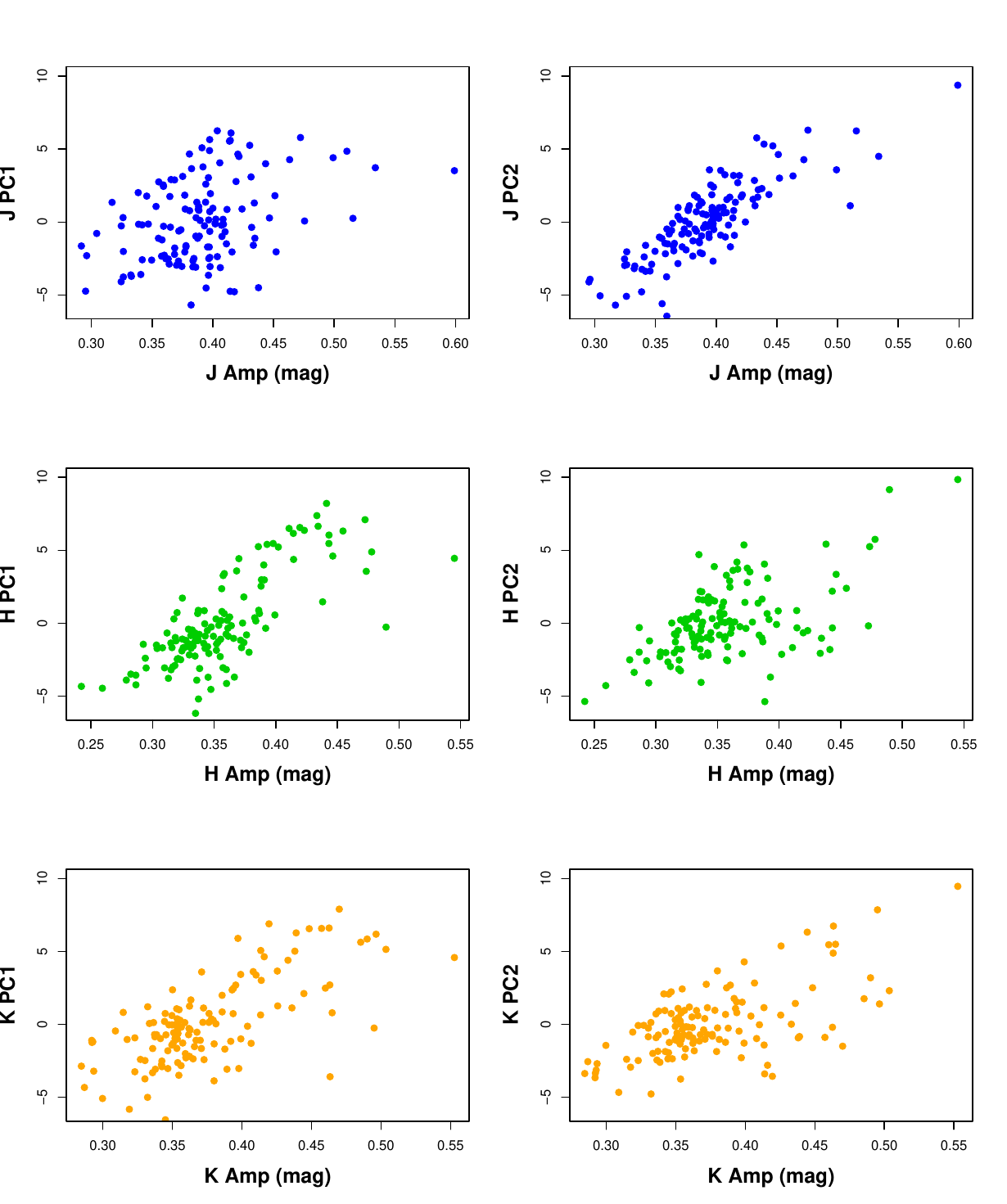}\\
  \caption{Dependence of the LC PC1 and PC2 principal components
  on the Amplitude in J,H and K colors.}
  \label{JHKPhA}
\end{figure}

To show the characteristics of these dependencies, we displayed
in Figure \ref{JHKPhA} the relationship between the PC1, PC2
principal components  as a function of the LC amplitude in all of
the three colors.

The properties of these scatter plots are very similar. There is a critical amplitude at about 0.38 mag in H and K color; however, it is less evident in the J.

\begin{figure}
  \includegraphics[width=8cm]{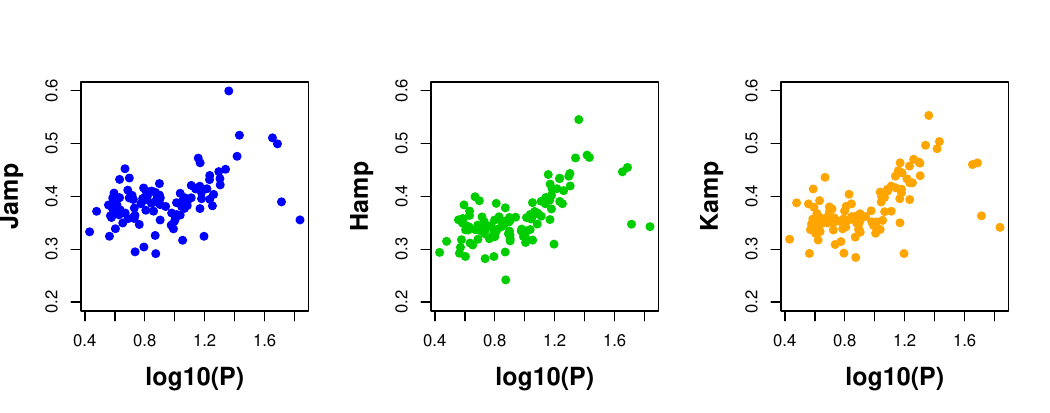}\\
  \caption{Period - Amplitude relationships in J,H and K color. The Figure shows clearly that
  the  $P=10^d$  period is  less discriminating in the amplitude
  J color than in H and K.}\label{JHKPAmp}
\end{figure}

The reason for this difference is clearly seen in Figure
\ref{JHKPAmp}. In the J color, the  $P=10^d$ period is much
less discriminative between low and high amplitude Cepheids than
in H and K. This fact can be seen in Table \ref{Jclus}. Group-1,
having the smallest mean magnitude in J color, has $P = 3.32^d$ and has
the smallest amplitude as medoid-1 demonstrates in Figure
\ref{Jmed}.

By contrast, the  $P=10^d$ period has a pronounced
discriminating power in the PCs in H and, in particular, in the K
color. In H color, the medoid of group-2 has the shortest period and
smallest amplitude; the smallest amplitude in K has 
smaller period than the $P=10^d$ in group-2. 

In the H and, in particular, in the K color, the LC medoids
characterizing  the groups of periods less than the cr$P=10^d$ value
have a flat plateau around the maximum brightness of the LC.
Passing the critical value in the direction of the greater values,
this property disappears, and the LCs become much sharped. 

 Figure \ref{JHKPAmp}  shows a tight relationship between the J,H,K amplitudes and the period. However, unlike Figure \ref{MJHK_P}, this relationship has a break at $P=10^d$. This tight connection also reveals a hidden stochastic variable responsible for it. Similar to the reasoning in the previous subsection, this can again be identified by the mass of the star.

Identifying the mass as the only essential physical variable for both absolute magnitude, period, and amplitude may raise the question of why we got two strong PCs instead of only one corresponding to this hidden stochastic variable. The answer is that the PCA is a linear theory, and the relationship between the amplitude and logP is nonlinear, so we need another PC to display it.

\subsection{Dependence of the PCs on the metallicity.}
\label{PCFeH}

Similarly to the period,  absolute magnitude, and amplitude of the LCs, we computed
Spearman correlations between PC1, PC2, and metallicity. The
results can be seen in Table~\ref{PCFe}.

Unlike the strong dependence of  PC1 and PC2 on the period,
the absolute magnitude and the amplitude of the first two PCs depend much less on the metallicity. One can infer from
Table~\ref{PCFe},  PC1, PC2 and PC3 have significant dependence on
$Fe/H]$ in the J color. In the other two colors, however, only PC3
has significant dependence on the metallicity in H and PC2 in the
K color.

\begin{figure}
  \includegraphics[width=8.1cm]{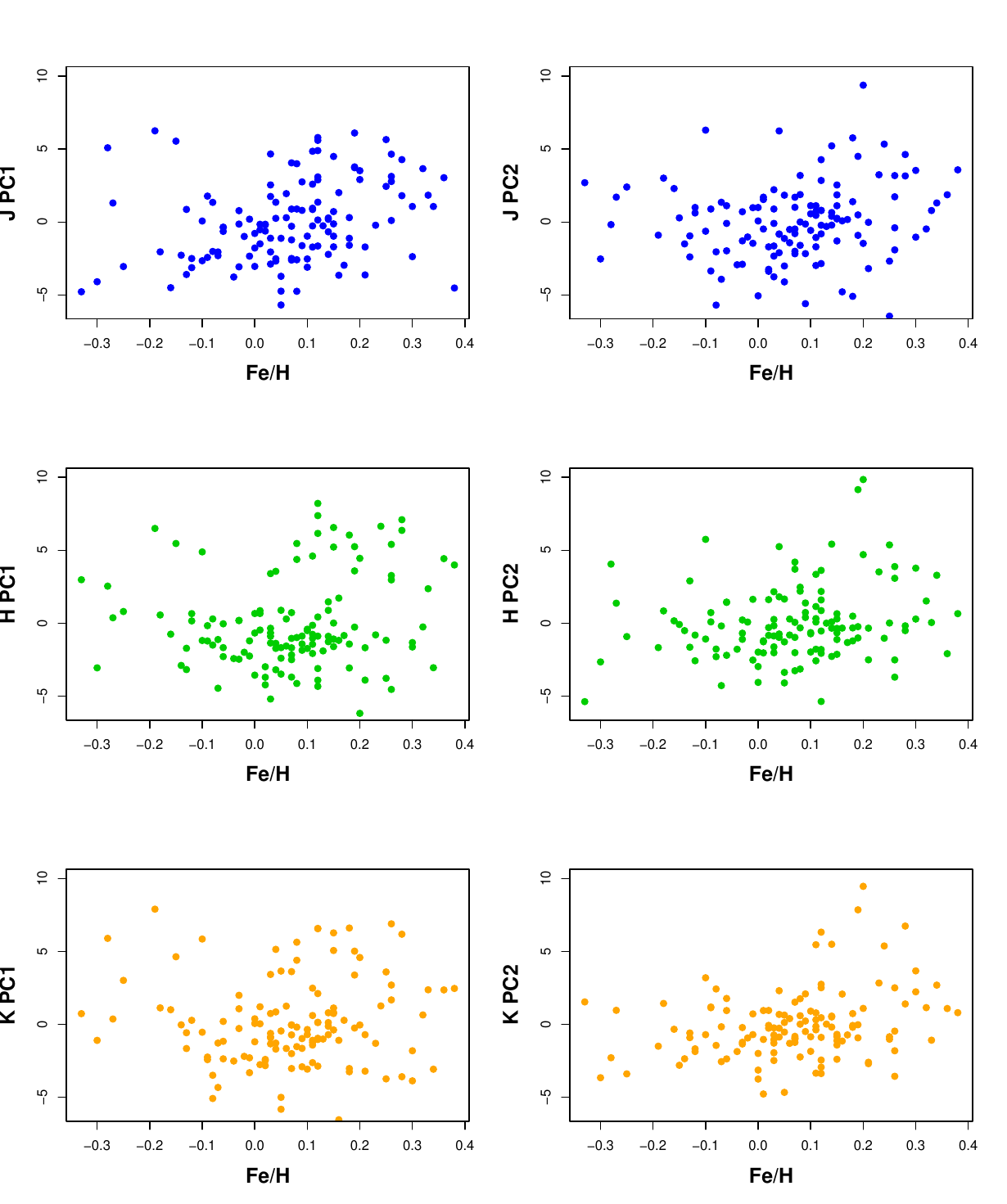}\\
  \caption{Dependence of the LC PC1 and PC2 principal components  on Fe/H in
  J,H and K colors. }\label{JHKFepc}
\end{figure}

As Figure \ref{JHKPhP} demonstrates,  the low value of PC1 and PC2
correspond to stars of $P < 10^d$ in all of the J,H,K colors.
A similar tendency may be identified in Figure \ref{JHKFepc} concerning the metallicity: low PC1 and PC2  correspond to lower [Fe/H], however,
with much less confidence. The general distributions of the
point patterns in Figure \ref{JHKFepc} seem more scattered.

From Table \ref{PCFe}, we can conclude that the LC in J color
depends significantly on the metallicity, but this dependence is
quite marginal in H and K colors. This accuracy is insufficient
to reliably estimate the metallicity from the NIR LC.

\begin{table}
  \centering
  \begin{tabular}{|c|c|c|c|c|c|c|}
    \hline
   Color    &    PC1     &    PC2     &    PC3     &   PC4    \\
  \hline
  J: p-val &  0.001 &  0.033 & 0.008 & 0.057  \\
  H: p-val &      0.164  &      0.053  &  0.004 & 0.485  \\
  K: p-val &      0.313  &  0.003 &      0.060  & 0.423  \\
  \hline
  \end{tabular}
  \caption{PC - Fe/H correlations obtained from Spearman's rank correlation,
  to avoid the possible effect of outliers. Significant correlations are marked
  with boldface.}\label{PCFe}.

\end{table}

\subsection{Connection of LC classes and physical parameters}

In Subsection \ref{pca}, we demonstrated that the observed
quantities representing the LCs are dominated by two strong PCs
describing 80\% and six ones about 90\% of the total variance of the LC brightness. We discussed their dependence on the period, absolute magnitude, amplitude, and metallicity in the following.

One may be interested in the distribution of these quantities
within the classes obtained in subsection \ref{medoid}. Comparing
these particular distributions within the clusters, one gets some information on the effect of the physical parameters on the LCs.

We compared the distribution within the different groups by 
the Kruscal-Wallis non-parametric tests \citep{Kru1952}. The
probabilities of the validity of the null hypothesis (i.e., the
differences are only by chance) are given in Tables \ref{Jclus},
\ref{Hclus} and \ref{Kclus} in the J,H, and K colors correspondingly. 

\begin{table}
	\centering
	\begin{tabular}{|c|c|c|c|c|c|}
		\hline
		Group & P (days) & MJ (mag) & Jamp (mag) & [Fe/H] \\
		\hline
		1 & 10.95 & -5.27 & 0.33 & 0.04 \\
		2 & 7.91  & -5.20 & 0.38 & 0.04 \\
		3 & 15.70 & -6.10 & 0.41 & 0.15 \\
		4 & 6.78  & -4.98 & 0.40 & 0.03 \\
		5 & 11.91 & -5.89 & 0.39 & 0.20 \\
		6 & 31.51 & -7.86 & 0.54 & 0.14 \\
		7 & 15.09 & -6.16 & 0.41 & 0.11 \\
		p-value &  < 1.0e-04 &  < 1.0e-04 &  < 1.0e-04 &  0.002 \\
		\hline
	\end{tabular}
	\caption{Mean values of physical quantities within groups in  J color obtained by the clustering procedure. The last row shows the probability that the differences are only by	chance. Significant statistical differences between groups are marked by boldface. The frequency of objects within each group is shown in		the bar chart of Fig. \ref{Jmed}.}
	\label{Jclus}
\end{table}

\begin{table}
	\centering
	\begin{tabular}{|c|c|c|c|c|c|}
		\hline
		Group & P (days) & MH (mag) & Hamp (mag) & [Fe/H] \\
		\hline
		1 & 14.52 & -6.51 & 0.38 & 0.15 \\
		2 & 5.90  & -5.20 & 0.32 & 0.04 \\
		3 & 9.53  & -5.82 & 0.34 & 0.08 \\
		4 & 6.37  & -5.12 & 0.36 & -0.01 \\
		5 & 18.03 & -6.88 & 0.45 & 0.05 \\
		6 & 23.34 & -7.00 & 0.44 & 0.17 \\
		7 & 16.82 & -6.30 & 0.37 & 0.12 \\
		p-value &  < 1.0e-04 & { < 1.0e-04} & { < 1.0e-04} & { 0.001} \\
		\hline
	\end{tabular}
	\caption{Mean values of physical quantities within groups in H color obtained by the
		clustering procedure. The last row shows the probability that the differences are only by
		chance. Significant statistical differences between groups marked by bold
		face. The frequency of objects within each group is shown in
		the bar chart of Fig. \ref{Hmed}.}
	\label{Hclus}
\end{table}

\begin{table}
	\centering
	\begin{tabular}{|c|c|c|c|c|c|}
		\hline
		Group & P (days) & MK (mag) & Kamp (mag) & [Fe/H] \\
		\hline
		1 & 6.27  & -5.50 & 0.35 & 0.04 \\
		2 & 8.63  & -5.36 & 0.35 & 0.04 \\
		3 & 15.83 & -6.24 & 0.42 & 0.10 \\
		4 & 8.19  & -5.59 & 0.36 & 0.08 \\
		5 & 12.79 & -5.96 & 0.37 & 0.08 \\
		6 & 20.63 & -7.11 & 0.46 & 0.15 \\
		7 & 11.51 & -6.07 & 0.39 & 0.10 \\
		p-value &  < 1.0e-04 &  < 1.0e-04 &  < 1.0e-04 & 0.126 \\
		\hline
	\end{tabular}
	\caption{Mean values of physical quantities within groups in  K color obtained by the
		clustering procedure. The last row shows the probability that the differences are only by
		chance. Significant statistical differences between groups marked by bold
		face. The frequency of objects within each group is shown in
		the bar chart of Fig. \ref{Kmed}.}\label{Kclus}
\end{table}

The tables clearly show that period, absolute magnitude, and amplitude discriminate
convincingly between the LC groups.  Nevertheless, it is also the case of the
metallicity in the J and H colors, with a much smaller significance. However, metallicity does not significantly discriminate between the obtained LC classes in the K color.

Inspecting Tables \ref{Jclus}, \ref{Hclus}, and \ref{Kclus} reveals
that group-6 has the longest period, greatest absolute brightness and amplitude, and highest metallicity. In contrast, the group having the
largest frequency of the objects (group-2 in J and H, group-1 in
K) has lower values in all of these variables.

\subsection{Discriminant power of physical parameters}
\label{lda}

In previous subsections, we compared distributions of some physical parameters (absolute magnitude, amplitude, period, metallicity) among the partitions of LCs obtained from their distributions in the parameter space defined by  PCs. 
The effectiveness of the classification depends on the physical parameters of observed LCs. It significantly depends on how accurately we know them and to what extent they affect the shape of the LCs. 

The different partitions may populate different areas in the parameter space. 
The strengths of separation between LC partitions in the parameter space can be characterized by the ratio of physical parameter variances between and within the partitions. We can search for direction in the parameter space along which the separation between the partitions is maximal. 

This task is performed by linear discriminant analysis (LDA).
If there are $g$ partitions, LDA results in $g-1$ discriminant functions, each is a linear combination of the original physical parameters. However, these do not necessarily describe significant separation. 

\begin{figure}
	\centering
	\includegraphics[width=8cm]{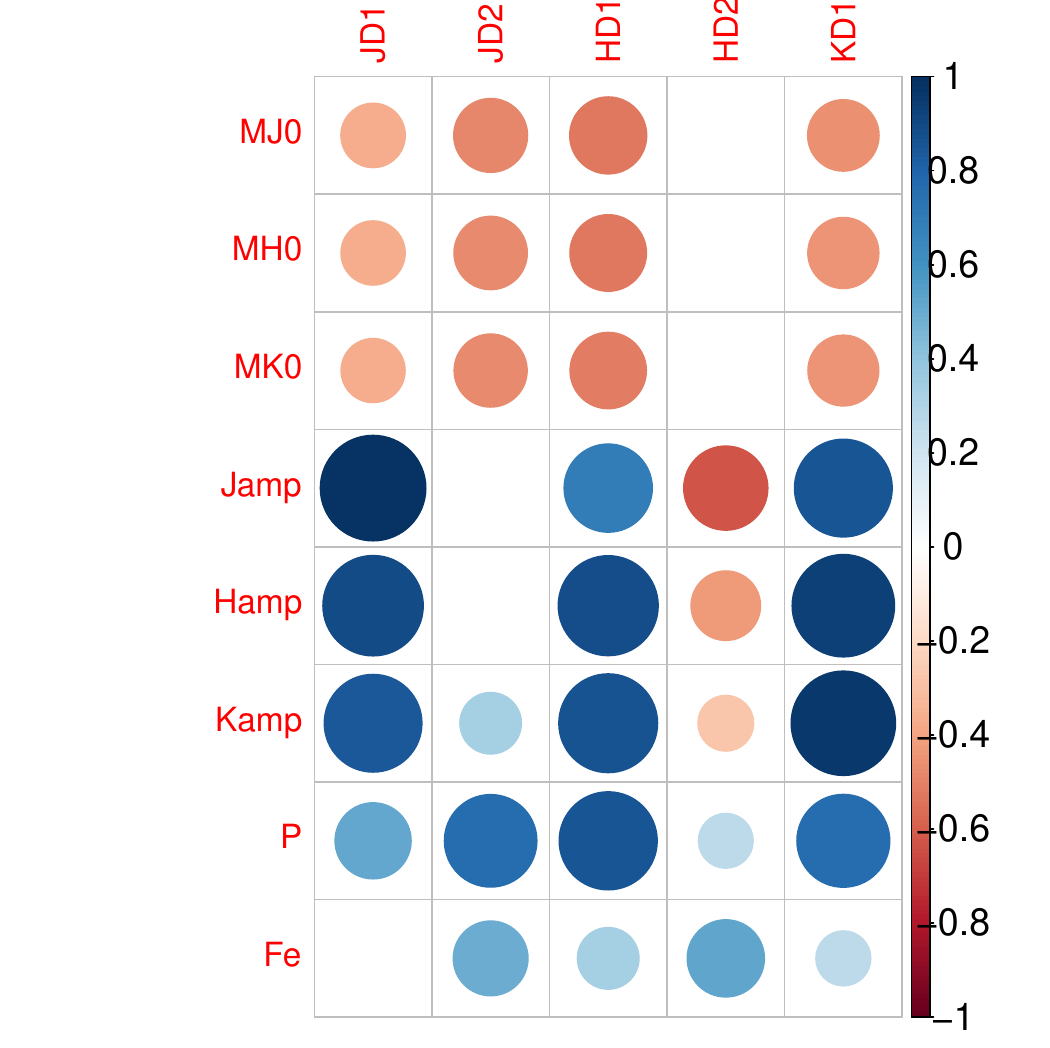}\\
	\caption{Colored display of the cross-correlations of LCs'  physical parameters with the significant discriminant functions in J,H,K colors obtained in the linear discriminant analysis (LDA). The size and tone of the circles indicate the strength of correlations. The blue color means positive, and the red color means negative correlations. The amplitudes have the highest and the metallicity the lowest correlations with each color's first (strongest) discriminant function.(The Figure was created using the \textcolor{red}{corrplot()} procedure of   R's $corrplot$ library.)}
	\label{coldis}
\end{figure}

We performed LDA analysis of the sample, consisting of physical parameters, using the \textcolor{red}{lda()} procedure in the $MASS$ library of R. We obtained two significant discriminant functions in colors J and H and one in color K, using Wilk's lambda test. Figure \ref{coldis} shows a color version of the cross-correlation (structure matrix) between the discriminant functions obtained and the original physical parameters. It can be seen from the figure that the LC amplitudes give the largest contribution while the metallicity makes the least.

It is worth noting that we also got two essential directions in the space of the PC parameters, which was explained by the dominant effect of the mass on the LCs of Cepheids. This is also likely to determine the number of significant discriminant functions.

Since these groups are different in all of the four observable
parameters (period, absolute magnitude, amplitude, metallicity) in the same way, we may
assume that there is a single, non-observable hidden statistical
variable responsible for these differences.

This statistical variable is not necessarily identical to some
of the physical variables responsible for the observational
characteristics of the LCs (e.g., mass). From a strictly theoretical
point of view, the mass and the metallicity are independent input
parameters of the models; still, metallicity cannot be deduced from the observation in this sample. \citep[For the relationship of mass to observational properties of Cepheids,
see the review paper of][and the references therein]{Cox1980}.

The statistical properties of the sample reflect the evolutionary
changes in star formation and metal enrichment in the given
Galactic environment, from which the sample was collected. This evolutionary process is reflected in the sampling of the objects and may statistically connect mass and metallicity in the sample.

For further details of the application of R, see the book \citep{Feigelson12} and references therein, and \citep{Zelko14} also recommended considering statistics and data mining with LSST in focus.

\subsection{Distinction between DCEP and DCEPS types}

All stars in our sample are  classical cepheids. They have the DCEP acronym in Berdnikov's catalogue \citep{Ber2000}. Nevertheless, Groenewegen's catalogue \citep{Gro2018} lists four stars in our sample as DCEPS. 

In GCVS (General Catalog of Variable Stars) DCEPS stars (also called as s-Cepheid) differ from typical DCEP variables with light amplitudes less than 0.5 mag in V color (0.7 in B). As a rule their optical LC is almost symmetrical. Their period is smaller than 7d, and they are probably first overtone pulsators \citep{Ant1990}.

\begin{figure}
	\centering
	\includegraphics[width=8cm]{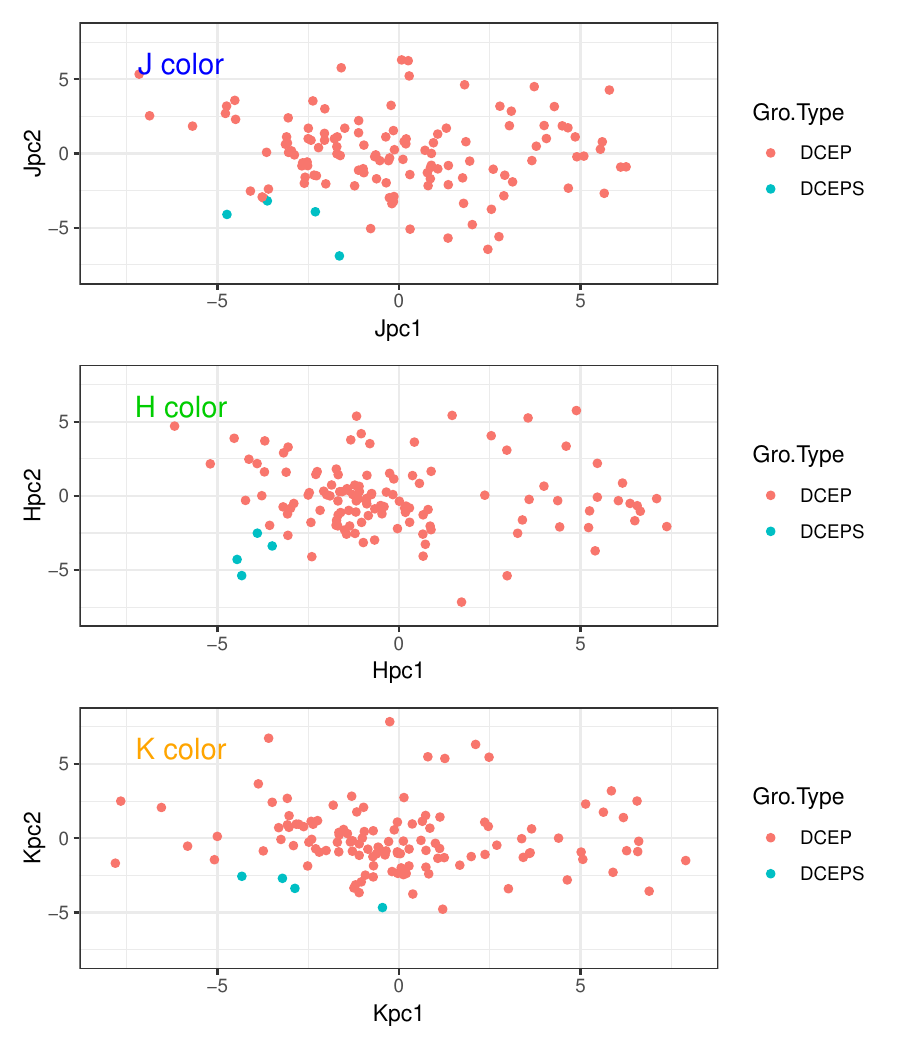}\\
	\caption{Scatterplot of PC1 and PC2 principal components in J,H and K colors. DCEPS stars are at the lower left edge of the DCEP distribution.}
	\label{ceptyp}
\end{figure}

We used PC1, PC2, ..., PC6 (significant PCs according to Figure \ref{eig}) to characterise the location of stars in PC's parameter space. Table \ref{xDCEP} shows the differences of coordinates' means in J,H,K colors. Seemingly, in their  DCEP, DCEPS' group means have the most significant difference in PC1 and PC2  coordinates. Figure \ref{ceptyp} shows J,H,K scatterplots of our sample stars in the (PC1,PC2) plain. 

\begin{figure}
	\centering
	\includegraphics[width=8cm]{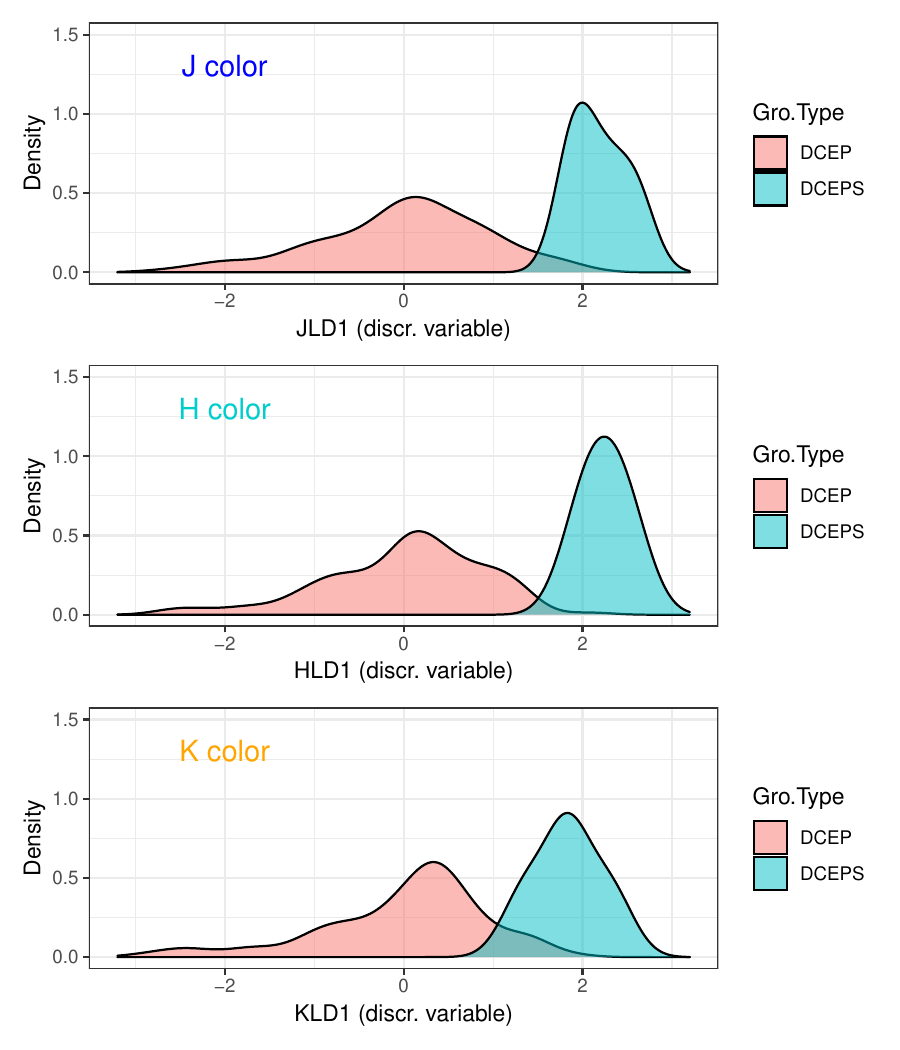}\\
	\caption{Probability density of DCEP and DCEPS stars along the best discriminating direction yielded by LDA in the PCs' parameter space.}
	\label{pld1}
\end{figure}

Similarly to subsection \ref{lda} we used LDA for estimating the best discriminating direction between DCEP and DCEPS stars' distribution in  PCs parameter space. As Figure \ref{pld1} clearly demonstrates there is a well defined difference in probability density function (likelihood functions assuming DCEP, DCEPS characteristics' are given) of these types along the best discriminating direction in the PC parameter space.

\begin{table}[ht]
\centering
\begin{tabular}{|r|r|r|r|r|r|r|}
  \hline
color & PC1 & PC2 & PC3 & PC4 & PC5 & PC6 \\ 
  \hline
J & 3.10 & 4.63 & 0.24 & -0.02 & 0.15 & 0.05 \\ 
  H & 4.11 & 3.96 & 0.21 & -0.25 & 0.14 & 0.26 \\ 
  K & 2.79 & 3.36 & 0.01 & -0.31 & -0.18 & -0.13 \\ 
   \hline
\end{tabular}
\caption{Difference in PC means of DCEP and DCEPS' groups in J,H and K colors. Seemingly, PC1 and PC2 components dominate the distance between DCEP and DCEPS groups in PC's parameter space, in all of the three colors. } 
\label{xDCEP}
\end{table}

\begin{figure}
	\centering
	\includegraphics[width=8cm]{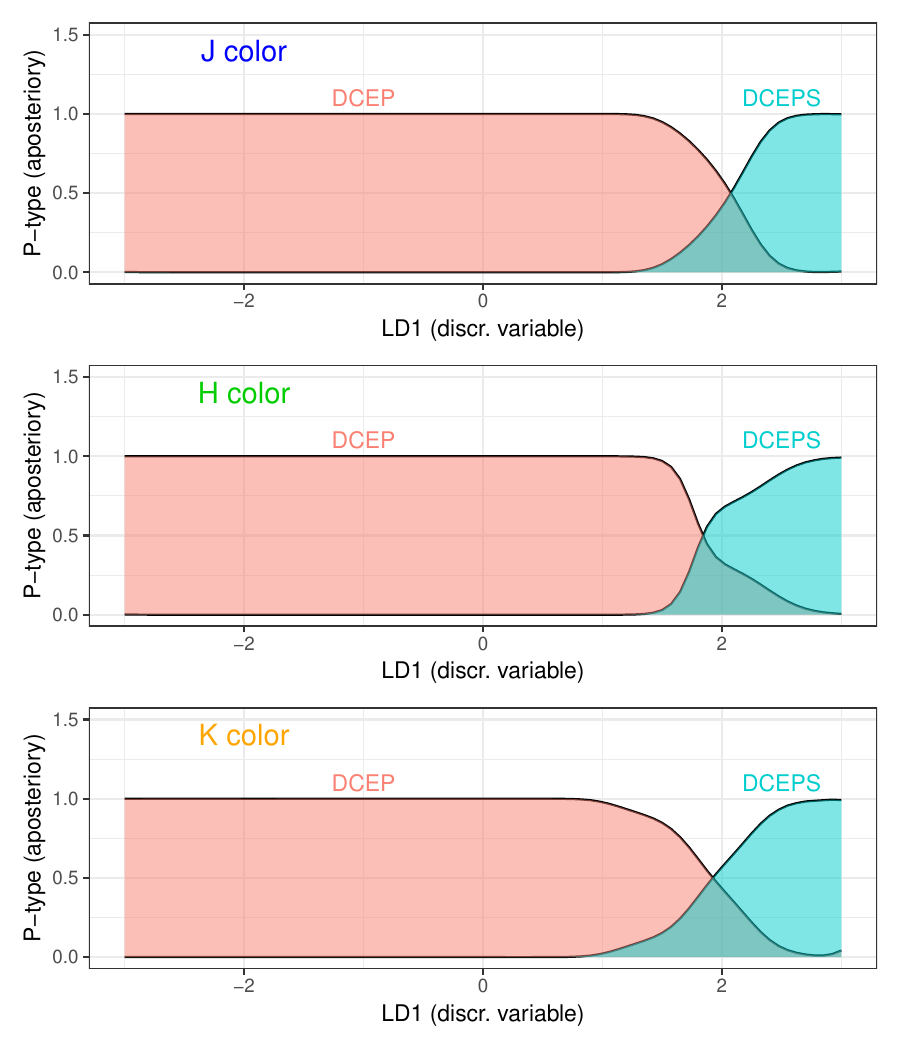}\\
	\caption{Aposteriori probability distribution of DCEP and DCEPS stars along the best discriminating direction in the PCs' parameter space. These probabilities can be used for classifying cepheids having known PCs but no DCEP or DCEPS types.}
	\label{apld1}
\end{figure}

According to Bayes theorem the $P(Type|\underline{PC})$ conditional probability of cepheid $Type$, assuming the $\underline{PC}$ vector, equals to $P(\underline{PC}|Type)$ likelihood multiplied by $P(Type)$ apriori probability of cepheid type and divided by $P(\underline{PC})$, the probability  of having $\underline{PC}$. The aposteriori probabilities obtained in this way (see Figure \ref{apld1})  can be used for classifying cepheid variables for which no DCEP, DCEPS types but only LC PCs are given.

\begin{equation}\label{Bay}
  P(Type|\underline{PC})=\frac{P(\underline{PC}|Type)P(Type)}{P(\underline{PC})}
\end{equation}

\section{Conclusions}

The parameters determining a star's static or dynamic behavior are to be extracted from observations, but theoretical model computations also need to do this. However, although our models were improved during the last half-century, they cannot represent all of the physical processes in pulsating stars and forming the observables, not considering those evolution phases that bring in random processes. So, the purely empirical and statistical approaches to analyze the observed features are fundamental. 

The metallicity of a star carries information about the environment in which it was formed and dredged up processes that happened during its evolution. Its value is generally determined from spectroscopic observation analyzing the shape of spectral lines. However, metal content also influences the transport processes in the stars and, through it, the spectral distribution of the radiated flux [see \citep{Gabor23a} and references therein]. Hence, the light curves observed by different filters depend also on it. However, the progression and regression of the ionization fronts strongly modulate the short wavelength transparency and transport. This gives rise to the asymmetry observed in the high energical and optical range, but at near-infrared, the outgoing flux is determined by the temperature of the zone from where photons can escape, which follows roughly the variation of the radius of the pulsating stars.

Therefore, inferring physical parameters from light curves is a legitimate endeavor that has yielded many results. 
As in all cases, when starting from empirical data, one must consider how the sampled data covers the parameter space, what biases are present in it, and what the errors of the individual measurements are. The statistical inferences we make from this - such as the PCA shown in a specific example above - help us to find and build the physical models that describe the systems we observed. The theoretical models thus built on these results may still include additional parameters that can only be estimated by comparing them to the measurements.

Obviously, if a sample is only partial, it is not sufficient in itself to provide exhaustive knowledge of the observed objects or systems. If we combine several samples and thus study a mixed one, we face another problem and the first task is to separate the mixed sample so that the identified groups contain homogeneous data. In large surveys, this classification problem is the first step.

Of course, the data can be represented mathematically in other ways that can help identify relationships whose physical causes are not known but are revealed by the structure of the data. By trial and error, we were able to find formulas for RR Lyare light curves that allowed us to approximate the metallicity from the Fourie series representation. This was then successfully implemented for other filters and also applied to Cepheids. However, the mathematical formulas obtained cannot be related to a physical understanding, and the errors, which depend on the accuracy of the measurements and the distribution of the samples, can only be estimated by additional calculations. 

Statistical methods provide consistent estimates from data samples and can be automated using direct procedures for millions of mixed samples. In the procedure described above, the light curves measured in each infrared band were analyzed independently without any prior assumptions. The similarity of the shape of the principal components to the Fourier series elements shows why this representation was successful. The dependence of the light curves measured in each band on metallicity differed and gave a hint that by treating them together by vectorizing them, we could achieve a level of detectability. This possibly reflects the fact that in the optical and shorter wavelength bands, the asymmetry of the light curves is dominant and carries more information. The work mentioned in the introduction and several others show that using the full range of available photometric data together significantly improves the accuracy of estimation. 

\section{Summary}
\label{summary}
We studied the relationship between the form of the LCs and the physical parameters of classical Cepheids in a NIR sample. To carry out this analysis,
we have performed a multivariate satisical study of 131 Cepheid LCs based on
their J,H,K colors by applying PCA. For this reason, we treated the LCs as cases in a multivariate data matrix. 

We utilized the observed data points as coordinates and represented the LCs in the 20D parameter space in each color separately. In this way, all LCs were
represented by a point in the 20D parameter space in every color. 

We characterized the similarity (distance) between two LCs by their $\chi^2$ distance. Performing a PCA among the variables of
the 20D parameter space and making use of the obtained PCs, we introduced an orthogonal coordinate system in this space and get
squared Euclidean distances between the LCs.

The set of eigenvalues obtained from the PCA allowed us to reduce the dimensions of the parameter space significantly
populated by the LCs in all three colors; finally, six eigenvalues differed significantly from the purely random case.

Making use of the six orthogonal eigenvectors belonging to the first six eigenvalues, we defined a 6-dimensional subspace in the
20D parameter space of the observed quantities and characterized the similarities (distances) between the individual LCs by squared Euclidean distances.

The introduced Euclidean distances between the LCs enabled us to get domains of similar objects in the parameter space. We used for this purpose the {\tt\textcolor{red}{pamk()}}  procedure of the {\it fpc} library of the  R statistical package.
We got  LC templates ({\it medoids}), and the \textcolor{red}{pamk()} procedure resulted in 7 groups in all the $J,H$, and $K$ colors.

As we demonstrated, the observed quantities representing the LCs are dominated by two strong PCs describing 80\% and the six
ones about 90\% of the total variance of the variables.  We discussed their dependence on the period, absolute magnitude, amplitude, and metallicity.

To reduce the effect of outliers, we computed Spearman rank correlations between the periods, absolute magnitudes, and the first six PCs in
each of the three colors.  The first three PCs have correlations at a very high level of significance. The Spearman correlations revealed significant relationships between
the amplitude and the first two PCs.

Unlike the strong dependence of  PC1 and PC2 on the period, absolute magnitude, and amplitude, the first two PCs depend on the metallicity with much less
significance. As Spearman correlations demonstrated, the PCs, the main characteristics of the LCs, depend only marginally on the metallicity of the sample objects
in these colors.

The shape of the LCs shows some symmetry concerning the maximum brightness in the H and K colors. It is not the case, however, in the J where the rise of the brightness is steeper than the decline.

The fact that PC1 and PC2 represent 80 percent of the total variance imposes a firm limit on the number of physical variables involved in the shape of the NIR LCs. Nevertheless, we identified the mass as the only dominant parameter. The two strong PCs appear since PCA is a linear theory, and the relationship between mass and other parameters is nonlinear. 

 Recently, time domain astronomy is getting increasing significance in many branches of astrophysics. We thought, an approach to solve the invers problem, i.e. getting physical parameters from LCs of classical cepheids has basic importance in many fields of astrophysics, including cosmology. Since interstellar extinction is the smallest in NIR spectral range, it gives a special significance.

Performing linear LDA on LC partitions obtained by grouping our sample stars in PC’s parameter space we estimated the significance of typical LC parameters in influencing the shape of observed LCs in J, H, K colors. As Figure 15. demonstrates clearly the most important parameter is LC’s amplitude. However, it is not a trivial issue, since LCs were normalized to the amplitude before performing PCA. 

Although, Berdnikov’s widely used catalogue assigned DCEP type to all our stars the catalog of Groenewegen listed four of our them as DCEPS. As mentioned above by combining LDA and Bayes theorem, we determined a way to classify DCEP, DCEPS types assigning aposteriory probabilities to the stars having one of these types.

In P > 10d time domain the maximum’s shape of our sample stars is more peaked than those having P < 10d period. Additionally, some of them have a sharp small peak at maximum intensity.  There are still no dynamical atmospheric models to theoretically explain this phenomenon.

We mentioned in subsection 5.4. that there is a difference in metallicity distribution between stars of P > 10d and P< 10d. We explained it in subsection 5.6. in the following way: “The statistical properties of the sample reflect the evolutionary changes in star formation and metal enrichment in the given Galactic environment, from which the sample was collected. This evolutionary process is reflected in the sampling of the objects and may statistically connect mass and metallicity in the sample.”. This is an interesting result, although does not relate to pulsation astrophysics.

\section{Acknowledgements}

We are indebted to Z. Koll\'ath, G. Kov\'acs, L. Szabados, and R. Szab\'o for
fruitful discussions and valuable suggestions. The support of NN129075 and K129249 NKFI grants and the Lend\"ulet Program of the Hungarian Academy of Sciences, project No. LP2018-7/2020 are acknowledged.

 This work has made use of data from the European Space Agency (ESA) mission
{\it Gaia} (\url{https://www.cosmos.esa.int/gaia}), processed by the {\it Gaia}
Data Processing and Analysis Consortium (DPAC,
\url{https://www.cosmos.esa.int/web/gaia/dpac/consortium}). Funding for the DPAC
has been provided by national institutions, in particular, the institutions
participating in the {\it Gaia} Multilateral Agreement.

\section{Data availability}

 The derived data generated in this research will be shared upon reasonable request to the corresponding author.

\bibliography{NIR_Cepheid}
\appendix
\section{List of R routines used in our paper}

The name of respective libraries (packages) are given in brackets. 

\begin{itemize}
    \item  {\tt corrplot} visualises a correlation matrix (corrplot)
    \item {\tt jackknife} makes Jackknife estimation (bootstrap)
    \item  {\tt lda} linear discriminant analysis (MASS)
    \item  {\tt pamk} partitions around medoids (fpc)
    \item  {\tt predict} predicts the values based on the input data (car)
    \item  {\tt princomp}  performs principal component analysis (stats)
    \item  {\tt smooth.spline} fits a cubic smoothing spline to the supplied data (stats)
\end{itemize}

Guide and necessary source files for installing R statistical language can be found at URL: https://www.r-project.org/. Currently, the CRAN package repository features 21254 available libraries (packages). 

(See URL: https://cran.r-project.org/web/packages/)

\label{lastpage}
 \end{document}